\documentclass[12pt]{article}
\usepackage{graphicx,color}
\usepackage{XoohmE}
\usepackage{booktabs}
\def\bo{{\raise.005ex\hbox{\large$\Box$}}}

 \parskip=\medskipamount                 
 \thicklines                         

 \def\e{\epsilon}
 \def\ve{\varepsilon}
 
 \def\der{\partial}
 \def\brr{\begin{equation}}
 \def\err{\end{equation}}
 \def\brr{\begin{eqnarray}}
 \def\err{\end{eqnarray}}
 \def\ba{\left(\begin{array}}
 \def\ea{\end{array}\right)}
 \def\lf{\left.\begin{array}{c}}
 \def\rf{\end{array}\right.}
 
 \def\dslash{\hbox{\ooalign{$\displaystyle\partial$\cr$/$}}}

 \newcommand{\dr}{\raise.3ex\hbox{$\stackrel{\leftarrow}{\partial }$}{}}
 \newcommand{\dl}{\raise.3ex\hbox{$\stackrel{\rightarrow}{\partial}$}{}}
 \newcommand{\topi}{\raise.3ex\hbox{$\stackrel{\pi}{\longrightarrow}$}{}}

 \def\qd{{\kern0.5pt
                   q \kern-5.05pt \raise5.8pt\hbox{$\textstyle.$}\kern 0.5pt}}

\def\4{\oplus}
\def\8{\otimes}

\def\spin{\mathfrak{spin}}

\definecolor{Hey}{rgb}{.9,.05,.4}
\definecolor{orange}{rgb}{1,.5,0}
\definecolor{plum}{rgb}{.4,0,.6}
\definecolor{R}{rgb}{1,0,0}
\definecolor{G}{rgb}{0,1,0}
\definecolor{B}{rgb}{0,0,1}
\font\cfnt=lcircle10 at 9pt
\def\lplus{\mathop{\kern2pt
            \raise1.275ex\hbox to0pt{\cfnt\char"07\hss}\kern-.6pt+}}
\def\YT#1#2{\vcenter{\hbox{\vbox{\baselineskip0pt\parskip=\medskipamount%
             \def\B{$\sqcap$\llap{$\sqcup$}\kern-1.9pt}
              \def\Bd{\hbox{\kern2.4pt\raise.4pt\hbox{$\cdot$}\kern-5.7pt\B\kern0pt}}
              \def\4{\raise.25pt\hbox to0pt{\hss\kern2pt--\hss}}
              \def\Z{\hfil\vskip-5.9pt}\lineskiplimit0pt\lineskip0pt%
               \setbox0=\hbox{#1}\hsize\wd0\parindent=0pt#2}\,}}}
\newdimen\parshift\parshift=\parindent
\catcode`@=11
 \long\def\@footnotetext#1{\insert\footins{\reset@font\footnotesize\interlinepenalty%
  \interfootnotelinepenalty\splittopskip\footnotesep\splitmaxdepth\dp\strutbox%
   \floatingpenalty\@MM\hsize\columnwidth\addtolength{\hsize}{-2\parindent}
    \@parboxrestore\protected@edef\@currentlabel{\csname p@footnote\endcsname\@thefnmark}
      \color@begingroup
       \@makefntext{\rule\z@\footnotesep\ignorespaces#1\@finalstrut\strutbox}
        \color@endgroup}}
 \long\def\@makefntext#1{\hglue\parshift
                         \vbox{\noindent\hb@xt@0em{\hss\@makefnmark\,}#1}}
\catcode`@=12
 \def\ba{\left(\begin{array}}
 \def\ea{\end{array}\right)}
 
 \def\der{\partial}
 \def\brr{\begin{eqnarray}}
 \def\err{\end{eqnarray}}

 \def\S{\mathbb S}

 \def\dslash{\hbox{\ooalign{$\displaystyle\partial$\cr$/$}}}
 \newcommand{\fr}[2]{{\textstyle\frac{#1}{#2}}}

 \def\blue{\color{blue}}
 \def\red{\color{red}}
 \def\green{\color{green}}
 \def\purple{\color{plum}}
 \HarvTitles 
 \seceq

\def\be{\begin{equation}}
\def\ee{\end{equation}}
\def\bea{\begin{eqnarray}}
\def\eea{\end{eqnarray}}

 \renewcommand{\theequation}{\arabic{section}.\arabic{equation}}

 \def\qd{{\kern0.5pt
                   q \kern-5.05pt \raise5.8pt\hbox{$\textstyle.$}\kern 0.5pt}}

\begin{document}

 \thispagestyle{empty}

 {\hbox to\hsize{July 2009 \hfill {SUNY-O/701}}}

 \vspace{.5in}

 \begin{center}
 {\Large\bf Dimensional Enhancement via Supersymmetry}
 \\[.25in]
 {M.G.\,Faux$^a$, K.M.\,Iga$^b$, and G.D.\,Landweber$^c$}\\[3mm]
 {\small\it
  $^a$ Department of Physics,
      State University of New York, Oneonta, NY 13820\\[-1mm]
  {\tt  fauxmg@oneonta.edu}
  \\
  $^b$Natural Science Division,
      Pepperdine University, Malibu, CA 90263\\[-1mm]
  {\tt  Kevin.Iga@pepperdine.edu}
  \\
 $^c$Department of Mathematics, Bard College, Annandale-on-Hudson, NY
 12504-5000\\[-1mm]
  {\tt gregland@bard.edu}
 }\\[.8in]

  {\bf ABSTRACT}\\[.01in]
 \end{center}
 \begin{quotation}
 We explain how the representation theory
 associated with supersymmetry in diverse dimensions is
 encoded within the representation theory of supersymmetry in
 one time-like dimension.  This is enabled by algebraic criteria, derived, exhibited, and
 utilized in this paper, which indicate
 which subset of one-dimensional supersymmetric models describe ``shadows" of
 higher-dimensional models.  This formalism delineates that minority of one-dimensional
 supersymmetric models which can ``enhance" to accommodate
 extra dimensions.  As a consistency test, we use our formalism to reproduce
 well-known conclusions about supersymmetric field theories using
 one-dimensional reasoning exclusively.  And we introduce the notion
 of ``phantoms" which usefully accommodate higher-dimensional gauge
 invariance in the context of shadow multiplets in supersymmetric
 quantum mechanics.
 \end{quotation}

 \vspace{.8in}

 ${~~~}$ \newline PACS: 04.65.+e

 \pagebreak

 \section{Introduction}
 Supersymmetry \cite{SS} imposes increasingly rigid constraints on the
 construction of quantum field theories \cite{QFT} as the number of spacetime
 dimensions increases. Thus, there are fewer supersymmetric models in
 six dimensions than there are in four, and fewer yet in
 ten dimensions \cite{Diverse}. In eleven dimensions there seems to be a unique
 possibility \cite{CJS}, at least on-shell.\footnote{Anomaly freedom imposes seemingly distinct algebraic
 constraints which make this situation even more interesting.}
 However, the off-shell representation theory
 for supersymmetry is well understood only for relatively few
 supersymmetries, and remains a mysterious subject in contexts
 of special interest, such as $N=4$ Super Yang Mills theory,
 and the four ten-dimensional supergravity theories \cite{GSW}.

 Many lower-dimensional models can be obtained from
 higher-dimensional models by dimensional reduction \cite{KK}.  Thus, a
 subset of lower-dimensional
 supersymmetric theories derives from the landscape of possible ways that extra
 dimensions can be removed.  But most
 lower-dimensional theories do not seem to be obtainable from higher
 dimensional theories by such a process; they seem to exist only in
 lower dimensions.  We refer to a lower-dimensional model obtained
 by dimensional reduction of a higher-dimensional model as the
 ``shadow" of the higher-dimensional model.  So we could re-phrase
 our comment above by saying that not all lower dimensional
 supersymmetric theories may be interpreted as shadows.

 It is a straightforward process to construct a shadow
 theory from a given higher dimensional theory.  But it is a more subtle
 proposition to construct a higher-dimensional supersymmetric model from a
 lower-dimensional model, or to determine whether a lower dimensional model
 actually does describe a shadow, especially of a higher dimensional
 theory which is also supersymmetric.  We have found resident
 within lower-dimensional supersymmetry an algebraic key which
 provides access to this information.  A primary purpose of this paper is
 to explain this.

 It is especially interesting to consider reduction to one time-like
 dimension, by switching off the dependence of all fields on all of the
 spatial coordinates.  Such a process reduces quantum
 field theory to quantum mechanics.  Upon making
 such a reduction, information regarding the spin representation
 content of the component fields is replaced with ${\cal R}$-charge
 assignments.  But it is not obvious whether the full higher-dimensional
 field content, or the fact that the one-dimensional model can be
 obtained in this way, is accessible information given the one-dimensional theory
 alone.  As it turns out, this information lies encoded within the extended one-dimensional supersymmetry transformation
 rules.

 We refer to the process of re-structuring a one-dimensional theory so that fields
 depend also on extra dimensions in a way consistent with covariant $\spin(1,D-1)$
 assignments and other structures, such as higher-dimensional
 supersymmetries, as ``dimensional enhancement".  This
 process describes the reverse of dimensional reduction.  We like to
 envision this in terms of the relationship between a
 higher-dimensional ``ambient" theory, and the restriction to a
 zero-brane embedded in the larger space.  A supersymmetric quantum
 mechanics then describes the ``worldline" physics on the
 zero-brane.  And the question as to whether this worldline physics
 ``enhances" to an ambient spacetime field theory is the reverse of
 viewing the worldline physics as the restriction of a target-space
 theory to the zero-brane.

 If the particular supersymmetric quantum mechanics obtained by restriction of a given theory to a zero-brane
 depended on the particular $\spin(1,D-1)$-frame described by that
 zero-brane, the higher-dimensional theory would
 not respect $\spin(1,D-1)$-invariance.
 Thus, if a one-dimensional theory enhances into a
 $\spin(1,D-1)$-invariant higher-dimensional theory, then the higher-dimensional theory obtained in
 this way should be
 agnostic regarding the presence or absence of an actual zero-brane on which
 such a one-dimensional theory might live.  This observation, in conjunction with the
 requirement of higher-dimensional supersymmetry, provides the
 requisite constraint needed to resolve the enhancement question.  In
 particular, by imposing $\spin(1,D-1)$-invariance on the enhanced
 supercharge operator, we are able to complete the ambient
 field-theoretic supercharge operator given merely the ``time-like" restrictions
 of this operator.  We find this interesting and surprising.

 The proposition that one can systematically delineate those one-dimensional
 theories which can enhance to higher-dimensions, and also discern how the
 higher-dimensional spin structures may be switched back on, is empowered by the
 fact that the representation theory of one-dimensional
 supersymmetry is relatively tame when compared with the
 representation theory of higher-dimensional supersymmetry, for a
 variety of reasons.  This enables the prospect of disconnecting the
 problem of spin assignments from the problem of classifying and
 enumerating supersymmetry representations, allowing these concerns
 to be addressed separately, and then merged together afterwards.
 With this motivation in mind, we have been
 developing a mathematical context for the representation theory of
 one-dimensional supersymmetry, also with other collaborators.

 In a sequence of papers \cite{FG1,DFGHIL01,Prepotentials,Counter,RETM,Zeeman,Frames}, we have explored the connection between
 representations of supersymmetry and aspects of graph theory.  We have shown that elements of
 a wide and physically relevant class of one-dimensional supermultiplets with vanishing central charge are
 equivalent to
 specific bipartite graphs which we call Adinkras; all of the salient algebraic
 features of the multiplets translate into restrictive and defining features of
 these objects.  A systematic enumeration of those graphs
 meeting the requisite criteria would thereby supply means for a corresponding
 enumeration of representations of supersymmetry.

 In \cite{AT1,HDS}, we have developed the paradigm further,
 explaining how, in the case of $N$-extended
 supersymmetry, the topology of all connected Adinkras are specified
 by quotients of $N$-dimensional cubes, and how the
 quotient groups are equivalent to doubly-even linear binary block
 codes.  Thus, the classification of connected Adinkras is related to
 the classification of such codes. In this way we have discovered an interesting connection between
 supersymmetry representation theory and coding theory \cite{Codes}.  All of this
 is part of an active endeavor aimed at delineating a
 mathematically-rigorous representation theory in
 one-dimension.

 In this paper we use the language of Adinkras, in
 a way which does not presuppose a deep familiarity with this topic.
 We have included Appendix \ref{adinkrastuff} as a
 brief and superficial primer, which should enable the reader
 to appreciate the entirety of this paper self-consistently.
 Further information can be had by consulting our earlier papers on
 the subject.

 In this paper we focus on the special case of
 enhancement of one-dimensional $N=4$ supersymmetric theories into
 four-dimensional $N=1$ theories.  This is done to keep our
 discussion concise and concrete.  Another motivating reason is because
 the supersymmetry representation theory for 4D $N=1$ theories is well known.  Thus, part and
 parcel of our discussion amounts to a consistency check on the very
 formalism we are developing.  From this point of view, this paper
 provides a first step in what we hope is a continuing process by
 which yet-unknown aspects of off-shell supersymmetry can be
 discerned. In the context of 4D theories, we use standard physics
 nomenclature, and refer to $\spin(1,3)$-invariance as ``Lorentz"
  invariance.

  We should mention that the prospect that aspects of higher-dimensional supersymmetry might
 be encoded in one-dimensional theories was suggested years ago in
 unpublished work \cite{enuf} by Gates et al.  Accordingly, we had
 used that attractive proposition as a prime motivator for developing the
 Adinkra technology in our earlier work.  This paper represents a
 tangible realization of that conjecture.
 Complementary approaches towards resolving a supersymmetry representation
 theory have been developed in \cite{T1,T2,T3,T4,T5,T6,T7}. Other ideas
 concerning the relevance of one-dimensional models to
 higher-dimensional physics were explored in \cite{FS1,FKS}.

 This paper is structured as follows.

 In Section 2 we describe an algebraic context for discussing
 supersymmetry tailored to the process of dimensional
 reduction to zero-branes and, vice-versa,
 to enhancing one-dimensional theories.  We explain how
 higher-dimensional spin structures can be accommodated into
 vector spaces spanned by the boson and fermion fields,
 and how the supercharges can be written as first-order
 linear differential operators which are also matrices which act
 on these vector spaces.  This is done by codifying the supercharge
 in terms of diophantine ``linkage matrices", which describe the
 central algebraic entities for analyzing the enhancement question.

 In Section 3 we explain how Lorentz invariance allows one to
 determine ``space-like" linking matrices from the ``time-like"
 linking matrices associated with one-dimensional supermultiplets,
 and thereby construct a postulate enhancement.  We then use this to
 derive non-gauge enhancement conditions, which provide an important
 sieve which identifies those one-dimensional multiplets which
 cannot enhance to four-dimensional non-gauge matter multiplets.

 In Section 4 we apply our formalism in a methodical and pedestrian
 manner to the context of minimal one-dimensional $N=4$
 supermultiplets, and show explicitly how the known structure of
 4D $N=1$ non-gauge matter may be systematically determined using
 one-dimensional reasoning coupled only with a choice of 4D spin
 structure.  We explain also how our non-gauge enhancement condition
 provides the algebraic context which properly delineates the Chiral
 multiplet shadow from its 1D ``twisted" analog, explaining why the
 latter cannot enhance.

 In Section 5 we generalize our discussion to include
 2-form field strengths subject to Bianchi identities.  This allows
 access to the question of enhancement to Vector multiplets.  In the process we introduce the
 notion of one-dimensional ``phantom" fields which prove useful in
 understanding how gauge invariance manifests
 on shadow theories.  We use the context of the 4D $N=1$ Abelian
 Vector multiplet as an archetype for future generalizations.

 We also include five appendices which are an important part of this
 paper.
 Appendix A is especially important, as this provides the
 mathematical proof that imposing Lorentz invariance of the
 postulate linkage matrices allows one to correlate the entire
 higher-dimensional supercharge with its time-like restriction.
 We also derive in this Appendix algebraic identities related to
 the spin structure of enhanced component fields, which
 should provide for interesting study in the future generalizations
 of this work.

 Appendix B is a brief summary of our Adinkra
 conventions, explaining
 technicalities, such as sign conventions, appearing in the bulk of the paper.
 Appendix C explains the dimensional
 reduction of the 4D $N=1$ Chiral multiplet, complementary to
  the non-gauge enhancement program described
 in Section 4.
 Appendix D explains the dimensional
 reduction of the 4D $N=1$ Maxwell field strength multiplet,
 complementary to Section 5.  This shows in detail how
 phantom sectors correlate with gauge aspects of the
 higher-dimensional theory.
 Appendix E is a discussion of four-dimensional spinors useful
 for understanding details of our calculations.

 We use below some specialized terminology.
 Accordingly, we finish this introduction section by providing the following three-term
 glossary, for reference purposes:

 \noindent
 {\it Shadow}: We refer to the one-dimensional multiplet which results
 from dimensional reduction of a higher-dimensional multiplet as the
 ``shadow" of that higher-dimensional construction.

 \noindent
 {\it Adinkra}:  The term Adinkra refers to 1D supermultiplets
 represented graphically, as explained in Appendix
 \ref{adinkrastuff}.  We sometimes use the terms Adinkra,
 supermultiplet, and multiplet synonymously.

 \noindent
 {\it Valise}:  A Valise supermultiplet, or a Valise Adinkra, is one in which
 the component fields span exactly two distinct engineering
 dimensions.  These multiplets form representative elements of
 larger ``families" of supermultiplets derived from these using
 vertex-raising operations, as explained below.  Thus, larger
 families of multiplets may be unpacked, as from a suitcase (or a valise),
 starting from one of these multiplets

 \section{Ambient versus Shadow Supersymmetry}
 It is easy to derive a one-dimensional
 theory by dimensionally reducing any given higher-dimensional
 supersymmetric theory.
 Practically, this is done by switching off the dependence of all fields
 on the spatial coordinates, by setting $\der_a\to 0$. One way to envision this process
 is as a compactification, whereby the spatial dimensions are rendered compact
 and then shrunk to zero size.  Alternatively, we may envision this process as
 describing a restriction of a theory onto a zero-brane, which is a time-like
 one-dimensional sub-manifold embedded in a larger, ambient,
 space-time.  Using this latter metaphor, we refer to the restricted theory as the ``shadow" of the
 ambient theory, motivated by the fact that physical shadows are constrained
 to move upon a wall or a wire upon which the shadow is cast.

 \subsection{Ambient Supersymmetry}
 Supersymmetry transformation rules can be written in terms
 of off-shell degrees of freedom, by expressing all
 fields and parameters in terms of
 individual tensor or spinor components.   Thus, without loss of
 generality, we can write the set of boson components as
 $\phi_i$ and the set of fermion components as
 $\psi_{\hat{\imath}}$, without being explicitly committal as the
 the $\spin(1,D-1)$-representation implied by these index
 structures. Generically, a $\spin(1,D-1)$-transformation acts on
 these components as
 \brr \delta_L\,\phi_i &=&
      \fr12\,\theta^{\mu\nu}\,(\,T_{\mu\nu}\,)_i\,^j\,\phi_j
      \nonumber\\[.1in]
      \delta_L\,\psi_{\hat{\imath}} &=&
      \fr12\,\theta^{\mu\nu}\,(\,\tilde{T}_{\mu\nu}\,)_{\hat{\imath}}\,^{\hat{\jmath}}\,
      \psi_{\hat{\jmath}} \,,
 \label{comreps}\err
 where the label $L$ is a mnemonic which specifies these as
 ``Lorentz" transformations.
 Here, $(\,T_{\mu\nu}\,)_i\,^j$ represents the spin algebra as realized on the
 boson fields and
 $(\,\tilde{T}_{\mu\nu}\,)_{\hat{\imath}}\,^{\hat{\jmath}}$
 represents the spin algebra as realized on the fermion fields,
 while $\theta^{0a}$ parameterizes a boost in the $a$-th spatial
 direction and $\theta^{ab}$ parameterizes a rotation in the
 $ab$-plane.  According to the spin-statistics theorem,
 $(\,\tilde{T}_{\mu\nu}\,)_{\hat{\imath}}\,^{\hat{\jmath}}$ should
 describe a spinor representation and $(\,T_{\mu\nu}\,)_i\,^j$ should
 describe a direct product of tensor representations.  The spin representations
 may also involve constraints.  For example, boson components may configure as closed $p$-forms.

 In four-dimensions the $N=1$ supersymmetry algebra is generated by
 a Majorana spinor supercharge with components ${\cal Q}_A$ subject to the
 anticommutator relationship $\{\,{\cal Q}_A\,,\,{\cal Q}_B\,\}=2\,i\,G^\mu_{AB}\,\der_\mu$
 where $G^\mu_{AB}=-(\,\Gamma^\mu\,C^{-1}\,)_{AB}$.
 A parameter-dependent supersymmetry transformation
 is generated by $\delta_Q(\e)=-i\,\bar{\e}^A\,{\cal Q}_A$, where $\e_A$
 describes an infinitesimal Majorana spinor parameter,
 and $\bar{\e}^A=(\,\e^\dagger\,\Gamma_0\,)^A$
 is the corresponding barred spinor.  It proves helpful, for our express purpose of
 restricting to a zero-brane, to use a Majorana basis where
 all spinor components, and all four Gamma matrices, are real.\footnote{
 See Appendix \ref{spinbases} for specifics related to this
 basis.}
 Furthermore, in this basis, we have the nice result $G^0_{AB}=\delta_{AB}$.
 With this choice, we can re-write our
 supersymmetry transformation as $\delta_Q(\e)=-i\,\e^A\,Q_A$,
 where $Q_A=(\,\Gamma_0\,)_A\,^B\,{\cal Q}_B$.\footnote{This merely technical re-organization
 facilitates dimensional reduction of 4D multiplets, as done in
 Appendices \ref{Chishadow} and \ref{maxlinks}.}
 The four-dimensional
 $N=1$ supersymmetry algebra, written in terms of the
 operators $Q_A$, is then given by
 \brr \{\,Q_A\,,\,Q_B\,\} &=&
      2\,i\,\delta_{AB}\,\der_\tau
      -2\,i\,G^a_{AB}\,\der_a \,,
 \label{s1}\err
 where $x^0:=\tau$ is the time-like coordinate parameterizing the
 the zero-brane to which we intend to restrict, and $x^a:=(\,x^1\,,\,x^2\,,\,x^3\,)$ are the three space-like
 coordinates transverse to the zero-brane.  To dimensionally reduce a four-dimensional field
 theory to a one-dimensional field theory, we set
 $\der_a=0$.  In this way, the second term on the right-hand side of
 (\ref{s1}) disappears, and we obtain the one-dimensional
 $N=4$ supersymmetry algebra.

 It proves helpful to add a notational distinction, by writing
 $\delta_Q\,\phi_i=-i\,\e^A\,(\,Q_A\,)_i\,^{\hat{\imath}}\,\psi_{\hat{\imath}}$
 and
 $\delta_Q\,\psi_{\hat{\imath}}=-i\,\e^A\,(\,\tilde{Q}_A\,)_{\hat{\imath}}\,^i\,\phi_i$,
 appending a tilde to $\tilde{Q}_A$ when this describes a fermion transformation rule.
 The supercharges may be represented as first-order linear differential operators, as
 \brr (\,Q_A\,)_i\,^{\hat{\imath}} &=&
      (\,u_A\,)_i\,^{\hat{\imath}}
      +(\,\Delta^\mu_A\,)_i\,^{\hat{\imath}}\,\der_\mu
      \nonumber\\[.1in]
      (\,\tilde{Q}_A\,)_{\hat{\imath}}\,^i &=&
      i\,(\,\tilde{u}_A\,)_{\hat{\imath}}\,^i
      +i\,(\,\tilde{\Delta}^\mu_A\,)_{\hat{\imath}}\,^i\,\der_\mu
      \,,
 \label{ssm}\err
 where $u_A$, $\tilde{u}_A$, $\Delta^\mu_A$, and
 $\tilde{\Delta}^\mu_A$ are real valued ``linkage matrices" which play a
 central role in our discussion below.

 The matrices $(\,u_A\,)_i\,^{\hat{\imath}}$ describe ``links"
 corresponding to supersymmetry maps from the bosons $\phi_i$
 to fermions $\psi_{\hat{\imath}}$ having engineering
 dimension one-half unit greater than the bosons.
 Therefore, these
 codify ``upward"  maps connecting lower-weight fermions to
 higher-weight bosons.\footnote{The term ``weight" refers to the engineering dimension of the
 field.  We sometimes use the term weight in lieu of dimension, to avoid confusion with spacetime dimension.
 The weight of a field correlates with the vertex ``height" on an Adinkra diagram.}  Similarly, the matrices
 $(\,\tilde{u}_A\,)_{\hat{\imath}}\,^i$ codify ``upward" maps
 connecting lower-weight fermions to higher-weight bosons.  The matrices
 $(\,\Delta^\mu_A\,)_i\,^{\hat{\imath}}$ and
 $(\,\tilde{\Delta}^\mu_A\,)_{\hat{\imath}}\,^i$ codify ``downward"
 maps accompanied by their respective derivatives $\der_\mu$.

  The component fields may be construed so that the
 linkage matrices conform to a special structure, known as the Adinkraic structure.
 This says that there is at most one
 non-vanishing entry in each column and at most one non-vanishing entry
 in each row.  Moreover, the non-vanishing entries take the values
 $\pm 1$. All known higher-dimensional off-shell representations
 in the standard literature satisfy this condition.\footnote{The
 only counterexamples that we know of were contrived by us
 in \cite{Counter}, as special deformations of one-dimensional Adinkraic
 representations. And we suspect that these do not enhance. Further scrutiny
 will be needed to ascertain any relevance of non-Adinkraic multiplets to physics.
 We find it sensible for now to focus on Adinkraic representations, especially since all known
 field theoretic multiplets are in this class.}

 The supersymmetry algebra (\ref{s1}) implies
 \brr (\,u_{(A}\,\tilde{u}_{B)}\,)_i\,^j &=& 0
       \hspace{.4in}
       (\,\tilde{u}_{(A}\,u_{B)}\,)_{\hat{\imath}}\,^{\hat{\jmath}}
       \,\,=\,\, 0
       \nonumber\\[.1in]
       (\,\Delta_{(A}^{(\mu}\,\tilde{\Delta}_{B)}^{\nu)}\,)_i\,^j
       &=& 0
       \hspace{.4in}
       (\,\tilde{\Delta}_{(A}^{(\mu}\,\Delta_{B)}^{\nu)}\,)_{\hat{\imath}}\,^{\hat{\jmath}}
       \,\,=\,\, 0 \,,
 \label{fil1}\err
 which describes a higher-dimensional analog of the Adinkra
 loop parity rule described in \cite{FG1} and below, and also implies
 \brr (\,u_{(A}^{\phantom{\mu}}\,\tilde{\Delta}^\mu_{B)}
      +\Delta^\mu_{(A}\,\tilde{u}_{B)}^{\phantom{\mu}}\,)_i\,^j &=&
      \Lambda^\mu_{AB}\,\delta_i\,^j
      \nonumber\\[.1in]
      (\,\tilde{u}^{\phantom{\mu}}_{(A}\,\Delta^\mu_{B)}
      +\tilde{\Delta}^\mu_{(A}\,u_{B)}^{\phantom{\mu}}\,)_{\hat{\imath}}\,^{\hat{\jmath}}
      &=& \Lambda^\mu_{AB}\,\delta_{\hat{\imath}}\,^{\hat{\jmath}}
      \,,
 \label{fil2}\err
 where $\Lambda^\mu_{AB}=(\,\Gamma_0\,G^\mu\,\Gamma_0\,)_{AB}=(\,\Gamma_0\Gamma^\mu\Gamma_0\,C^{-1}\,)_{AB}$,
 whereby $\Lambda^0_{AB}=G^0_{AB}$ and $\Lambda^a_{AB}=-G^a_{AB}$.
 The equations (\ref{fil2}) play a central role in this
 paper.

 The classification of
 representations of supersymmetry in diverse dimensions is equivalent to the question of classifying and
 enumerating the possible sets of real linkage matrices which can
 satisfy the algebraic requirements in (\ref{fil1}) and
 (\ref{fil2}), and identifying the corresponding spin representation matrices
 $(\,T_{\mu\nu}\,)_i\,^j$ and
 $(\,\tilde{T}_{\mu\nu}\,)_{\hat{\imath}}\,^{\hat{\jmath}}$.

 \subsection{Shadow Supersymmetry}
 The one-dimensional $N=4$ superalgebra is specified by
 $\{\,Q_A\,,\,Q_B\,\}=2\,i\,\delta_{AB}\,\der_0$, which corresponds
 to (\ref{s1}) in the limit $\der_a\to 0$.   In this case, the
 supercharges are represented as
  \brr (\,Q_A\,)_i\,^{\hat{\imath}} &=&
      (\,u_A\,)_i\,^{\hat{\imath}}
      +(\,d_A\,)_i\,^{\hat{\imath}}\,\der_\tau
      \nonumber\\[.1in]
      (\,\tilde{Q}_A\,)_{\hat{\imath}}\,^i &=&
      i\,(\,\tilde{u}_A\,)_{\hat{\imath}}\,^i
      +i\,(\,\tilde{d}_A\,)_{\hat{\imath}}\,^i\,\der_\tau \,.
 \label{ss01}\err
 This is identical to (\ref{ssm}) except the index $\mu$ is
 restricted to the sole value $\mu=0$, and the down matrices have
 been re-named by writing $\Delta_A^0$ as
 $d_A$ and $\tilde{\Delta}_A^0$ as $\tilde{d}_A$.
 As mentioned above, the fields may be configured
 so that each linkage matrix has not more than one non-vanishing entry in each row and
 likewise in each column, and the non-vanishing entries are $\pm 1$.
 This specialized structuring enables the faithful translation of
 1D supercharges in terms of helpful and interesting graphs known as Adinkras, as mentioned
 in the Introduction.  The reader should consult Appendix \ref{adinkrastuff} for
 a simple-but-practical overview of this concept.

 The algebra obeyed by 1D linkage matrices may be
 obtained from (\ref{fil1}) and (\ref{fil2}) by allowing only the value 0 for the
 spacetime indices $\mu$ and $\nu$.  Thus, the linkage matrices are
 constrained by
 \brr (\,u_{(A}\,\tilde{u}_{B)}\,)_i\,^j &=& 0
       \hspace{.4in}
       (\,\tilde{u}_{(A}\,u_{B)}\,)_{\hat{\imath}}\,^{\hat{\jmath}}
       \,\,=\,\, 0
       \nonumber\\[.1in]
       (\,d_{(A}\,\tilde{d}_{B)}\,)_i\,^j
       &=& 0
       \hspace{.4in}
       (\,\tilde{d}_{(A}\,d_{B)}\,)_{\hat{\imath}}\,^{\hat{\jmath}}
       \,\,=\,\, 0 \,.
 \label{gram}\err
 These relationships imply a ``loop parity" rule, described in our earlier papers, which
 says that any closed bi-color loop on an Adinkra diagram must involve an odd number of edges
 with odd parity. The linkgage matrices are further constrained by
 \brr (\,u_{(A}\,\tilde{d}_{B)}
      +d_{(A}\,\tilde{u}_{B)}\,)_i\,^j &=&
      \delta_{AB}\,\delta_i\,^j
      \nonumber\\[.1in]
      (\,\tilde{u}_{(A}\,d_{B)}
      +\tilde{d}_{(A}\,u_{B)}\,)_{\hat{\imath}}\,^{\hat{\jmath}}
      &=& \delta_{AB}\,\delta_{\hat{\imath}}\,^{\hat{\jmath}} \,.
 \label{gral}\err
 In this context, the algebra defined by (\ref{gral}) was
 called a ``Garden algebra" by
 Gates et al., in \cite{GatesRana1,GatesRana2}, and the the matrices $u_A$ and $d_A$ were called
 Garden matrices.  The larger algebra given in (\ref{fil1})
 and (\ref{fil2}) generalizes
 this concept to diverse spacetime dimensions, and accordingly
 subsumes these smaller algebras.

 A one-dimensional supermultiplet is specified by the set
 of linkage matrices $u_A$, $\tilde{u}_A$, $d_A$, and $\tilde{d}_A$
 or equivalently by the Adinkra diagram representing these matrices.
 Given a set of linkage matrices one can construct the equivalent
 Adinkra.  Alternatively, given an Adinkra, one can use this to ``read off"
 the equivalent set of linkage
 matrices. Given either of these, one can ascertain supersymmetry
transformation rules
 and invariant action functionals from which one can study
 one-dimensional physics.  The linkage matrices associated with any
 Adinkra satisfy the algebra (\ref{gram}) and (\ref{gral})
 by definition.

 \section{Enhancement Criteria for Shadow Supermultiplets}
 The requirement that the linkage matrices appearing in the
 supercharges (\ref{ssm}) are $\spin(1,D-1)$-invariant has
 remarkable implications.  One of these is the fact that the
 ``space-like" linkage matrices $\Delta^a_A$ are completely determined by the
 ``time-like" linkage matrices $\Delta^0_A$.  The proof of this
 assertion is given as Appendix \ref{prlemma}, with the result
 \brr (\,\Delta^a_A\,)_i\,^{\hat{\imath}} &=&
      -(\,\Gamma^0\Gamma^a\,)_A\,^B\,(\,\Delta_B^0\,)_i\,^{\hat{\imath}}
      \nonumber\\[.1in]
      (\,\tilde{\Delta}^a_A\,)_{\hat{\imath}}\,^i &=&
      -(\,\Gamma^0\Gamma^a\,)_A\,^B\,(\,\tilde{\Delta}_B^0\,)_{\hat{\imath}}\,^i
      \,.
 \label{ga}\err
 It is interesting that the matrix $(\,\Gamma^0\Gamma^a\,)_A\,^B$ is precisely
 twice a boost operator in the $a$-th spatial direction, in the spinor representation.
 It is also interesting that the result (\ref{ga}) holds
 irrespective of the $\spin(1,D-1)$-representations described by the
 component fields.  That is, the assignment of the matrices
 $(\,T_{\mu\nu}\,)_i\,^j$ and
 $(\,\tilde{T}_{\mu\nu}\,)_{\hat{\imath}}\,^{\hat{\jmath}}$ defined
 in (\ref{comreps}) does not influence (\ref{ga}).  These
 nontrivial consequences are derived explicitly in Appendix
 \ref{prlemma}.\footnote{The Lorentz invariance of the linkage matrices does imply
 interesting and interlocking constraints on the allowable choices of
 $(\,T_{\mu\nu}\,)_i\,^j$ and
 $(\,\tilde{T}_{\mu\nu}\,)_{\hat{\imath}}\,^{\hat{\jmath}}$.
 These are exhibited in Appendix \ref{prlemma}.   Such correlations
 are certainly expected, and we suspect that
 equations (\ref{daa}) and (\ref{tcons}) have deep and useful
 implications, which we hope to explore in future work.}

 It is worth mentioning that the form of
 (\ref{ga}) agrees precisely with the linkage matrices derived from
 Salam-Strathdee superfields. Also, the appearance of $\Gamma^a$ on the
 right hand side is tied closely to the appearance of the $\Gamma^a$ in
 the defining supersymmetry algebra.

 The result (\ref{ga}) is the crux ingredient which allows one to determine
 whether a given one-dimensional supermultiplet describes the shadow
 of a higher-dimensional supermultiplet.  This follows because any
 one-dimensional multiplet organizes as (\ref{ssm})
 where the index $\mu$ assumes only the value 0.  To probe whether
 that multiplet describes a shadow, one creates ``provisional"
 off-brane linkages using the powerful expression (\ref{ga}).  Since there is
 no algebraic guarantee that the transformation rules so-extended
  will properly close the
 higher-dimensional superalgebra, nor that the boson and fermion
 vector spaces will properly assemble into representations of the
 higher-dimensional spin group, the higher-dimensional superalgebra itself, applied to this
 construction, provides the requisite analytic probe of that
 possibility:
 if the one-dimensional multiplet is a shadow then the
 provisional construction will close the higher-dimensional
 superalgebra; if it is not possible, then it will not.

 The supersymmety algebra in $D$-dimensions closes only if
 $(\,\Omega^\mu_{AB}\,)_i\,^j\,\der_\mu\phi_j=0$ and
 $(\,\tilde{\Omega}^\mu_{AB}\,)_{\hat{\imath}}\,^{\hat{\jmath}}\,\der_\mu\psi_{\hat{\jmath}}=0$,
 where we define the following useful matrices,
 \brr (\,\Omega^\mu_{AB}\,)_i\,^j &=&
      (\,u_{(A}^{\phantom{\mu}}\,\tilde{\Delta}^\mu_{B)}
      +\Delta^\mu_{(A}\,\tilde{u}_{B)}^{\phantom{\mu}}\,)_i\,^j
      -\Lambda^\mu_{AB}\,\delta_i\,^j
      \nonumber\\[.1in]
       (\,\tilde{\Omega}^\mu_{AB}\,)_{\hat{\imath}}\,^{\hat{\jmath}} &=&
      (\,\tilde{u}_{(A}^{\phantom{\mu}}\,\Delta^\mu_{B)}
      +\tilde{\Delta}^\mu_{(A}\,u_{B)}^{\phantom{\mu}}\,)_{\hat{\imath}}\,^{\hat{\jmath}}
      -\Lambda^\mu_{AB}\,\delta_{\hat{\imath}}\,^{\hat{\jmath}} \,.
 \label{matos}\err
 This requirement is a minor re-structuring of (\ref{fil2}).
 In this way, we have written the supersymmetry algebra as
 a linear algebra problem, cast as matrix equations.

 Many important supemultiplets exhibit gauge invariances, manifest
 as physical redundancies inherent in the vector
 spaces spanned by the component fields.  In these cases, the matrices $(\,\Omega^\mu_{AB}\,)_i\,^j$ and
 $(\,\tilde{\Omega}^\mu_{AB}\,)_{\tilde{\imath}}\,^{\tilde{\jmath}}$,
 are not unique.  Instead these describe classes of matrices
 interrelated by operations faithful to the gauge structure.
 We describe this interesting situation below, in Section \ref{phan}.
 It is useful, however, to begin our discussion with what we call
 non-gauge matter multiplets, which do not exhibit redundancies of
 this sort.  For this smaller but nevertheless interesting and relevant
 class of supermultiplets,
 the higher-dimensional supersymmetry algebra is satisfied only if
 \brr (\,\Omega^\mu_{AB}\,)_i\,^j &=& 0
      \nonumber\\[.1in]
      (\,\tilde{\Omega}^\mu_{AB}\,)_{\hat{\imath}}\,^{\hat{\jmath}}
       &=& 0 \,.
 \label{gregeq}\err
 We refer to these equations as our non-gauge enhancement
 criteria.  These enable a practical algorithm for testing whether a
 given 1D supermultiplet represents the shadow of a 4D non-gauge
 matter multiplet.

 We use the linkage matrices for a given 1D supermultiplet
 in conjuction with the 4D Gamma matrices to compute all of the
 $d\times d$ matrices $\Omega^\mu_{AB}$ and $\tilde{\Omega}_{AB}$,
 defined in (\ref{matos}).  If (\ref{gregeq})
 is satisfied, \ie, if all of these matrices are identically
 null, then the 1D multiplet passes an important, non-trivial, and
 necessary requirement for enhancement to a 4D non-gauge matter
 multiplet.  If these matrices do not vanish, then the 1D multiplet
 cannot enhance to a 4D non-gauge matter multiplet.  In the latter case,
 further analysis must be done to probe whether this multiplet can
 enhance to a gauge multiplets.  Equation (\ref{gregeq}) represents
 a useful ``sieve" in the separation of 1D multiplets into
 groups as shadows versus non-shadows.

 A second important sieve derives from the spin-statistics theorem.
 As it turns out, a minority of 1D multiplets actually pass the test
 (\ref{gregeq}).  But those that do come in pairs related in-part by
 a Klein flip, which is an involution under which the
 statistics of the fields are reversed --- boson fields are replaced
 with fermion fields and vice versa.  Thus, we can organize those multiplets
 which pass the test (\ref{gregeq}) into such pairs.  We then ascertain which elements of each
 pair satisfy the requirement that fermions assemble
 as spinors and the bosons as tensors.
 Those multiplets that do not pass this test describe another class of
 multiplets which do not describe ordinary shadows.  Typically, one
 multiplet out of each pairing satisfies the spin-statistics
 test while the other multiplet fails this test.\footnote{The important role of the Klein flip in the representation
 theory of superalgebras was addressed by one of the authors (G.L.) in
 previous work \cite{GL_Klein}.}

 In the explicit examples analyzed below in this paper, it is
 obvious when certain multiplets which pass the first enhancement
 test (\ref{gregeq}) fail the spin-statistics test.  This occurs
 when the multiplicity of fermions with a common engineering dimension
 is not a multiple of four, thereby obviating assemblage into 4D spinors.
 In fact our analysis below is remarkably clean.  In more general
 cases, we suspect that more careful attention to the implications
 of the Lorentz invariance of the provisional supercharge, codified
 by equations such as (\ref{daa}), will provide the requisite sophistication needed to
 address enhancement at higher $N$ and higher $D$.  We
 think this will be a most interesting undertaking.

 \section{Non-Gauge Matter Multiplets}
 In this section we impose our enhancement equation (\ref{gregeq}) on the linkage
 matrices associated with all of the minimal $N=4$ Adinkras, of which there are
 60 in total, to
 ascertain which of these represent shadows of 4D $N=1$ non-gauge matter multiplets.
 Since the represention theory for minimal irreducible multiplets in 4D $N=1$ supersymmetry
 is well known, this setting provides a natural laboratory
 for testing our technology.  The principal result of this section
 is that our enhancement equation properly corroborates
 what is known about non-gauge matter in 4D $N=1$ supersymmetry,
 thereby passing an important consistency test.  Another principal result of this
 section identifies our enhancement equation as the natural
 algebraic sieve which distinguishes the Chiral multiplet shadow from
 its ``twisted" analog.

 By the term ``non-gauge matter multiplets" we refer to 4D
 supermultiplets which involve component fields neither subject to gauge
 transformations nor subject to differential constraints,
 such as Bianchi identities.  This excludes the Vector and the Tensor multiplets,
 as well as the the corresponding field strength multiplets.
 We postpone a discussion of these interesting cases until the next
 section.  In fact, the only non-gauge matter multiplet in 4D $N=1$
 supersymmetry is the Chiral multiplet.\footnote{An Antichiral
 multiplet, which can be formed as the Hermitian conjugate of a Chiral multiplet,
 is not distinct from the latter as representation of the 4D $N=1$ supersymmetry algebra
 separate from inherent complex structures; the assignment
 of possible $U(1)$ charge assignments represents ``extra" data not considered overtly in this paper.  Ignoring the complex structure,
 the Chiral and the Antichiral multiplets have indistinguishable shadows.}
 As we will see, among the 60 different minimal $N=4$ Adinkras there are
 exactly four which satisfy our primary enhancement condition (\ref{gregeq}).  For two of
 these, the fermions configure as a spinor and the bosons configure
 as Lorentz scalars.  For the other two the fermions configure as
 Lorentz scalars while the bosons configure as a spinor.  The latter
 case fails the spin-statistics test, which says that fermions must
 assemble as spinors, and bosons must assemble as tensors.  Thus,
 our method identifies the two Adinkras which can provide shadows of
 4D minimal non-gauge matter multiplets.\footnote{Conceivably, the fact that
 there are two such enhanceable $N=4$ minimal Adinkras may relate
 to the fact that there are two complementary choices of complex
 structure, related to the Chiral and Antichiral multiplets, as mentioned in
 the previous footnote.}

 It is noteworthy that there are two separate minimal $N=4$ Adinkra families, related by a
 so-called twist, implemented by toggling the parity of one of
 the four edge colors.  Thus, the shadow of the Chiral multiplet has
 a twisted analog which cannot enhance to 4D.  That multiplet, which has been called
 the Twisted Chiral multiplet, describes 1D physics which cannot be obtained by restriction from
 four-dimensions.
 We have long wondered what algebraic feature
 distinguishes these two.\footnote{We learned about this interesting
 curiosity from Jim Gates, in the context of a former collaboration.}
 As it turns out, the linkage matrices for
 the Chiral multiplet shadow satisfy the enhancement equations
 (\ref{gregeq}) whereas the linkage matrices for the Twisted Chiral
 multiplet do not.  This answers this long-puzzling question.
 Details are presented below in this section.

 In order to ascertain whether a given Adinkra enhances to 4D we need to subject the
 corresponding linkage matrices to the space-like subset of the equations in
 (\ref{gregeq}).  (The time-like equations are
 satisfied automatically, since an Adinkra is a representation of 1D
 supersymmetry by construction.)
 In the case of testing enhancement to 4D $N=1$ supersymmetry, each of the two conditions in
 (\ref{gregeq}) describes 30 matrix equations for each of the 60 Adinkras to be tested, since for each of the
 three choices for $a$, the corresponding symmetric matrices $G^a_{AB}=G^a_{(AB)}$
 have ten independent components.  Thus, according to the crudest counting argument,
 in order to test both the bosonic and fermionic
 conditions in (\ref{gregeq}) for all the minimal $N=4$ Adinkras,
 we need to check $60\times 30\times 2=3600$ matrix equations, each involving
 products of $4\times 4$ matrices.  This is a simple matter which we have managed expediently using
 rudimentary Mathematica programming.\footnote{ More generally, testing
 enhancement to $D$ dimensions will involve at least $\fr12\,(\,D-1\,)\,d\,(d+1)$
 independent matrix equations per Adinkra, where $d$ is the
 number of fermions or bosons in the Adinkra. The minimal-size Adinkra in the
 case $D=1$, $N=16$ is 128+128, whereby $d=128$.  To ascertain whether one of these
 enhances to $D=10$, $N=1$ supersymmetry would involve
 $\fr12\,(9)\,(\,128\,)\,(\,129\,)=74,304$ equations, each involving
 products of $128\times 128$ matrices.  This number is
 not prohibitively large given contemporary computer resources.}

\begin{figure}
 \begin{center}
 \includegraphics[width=5in]{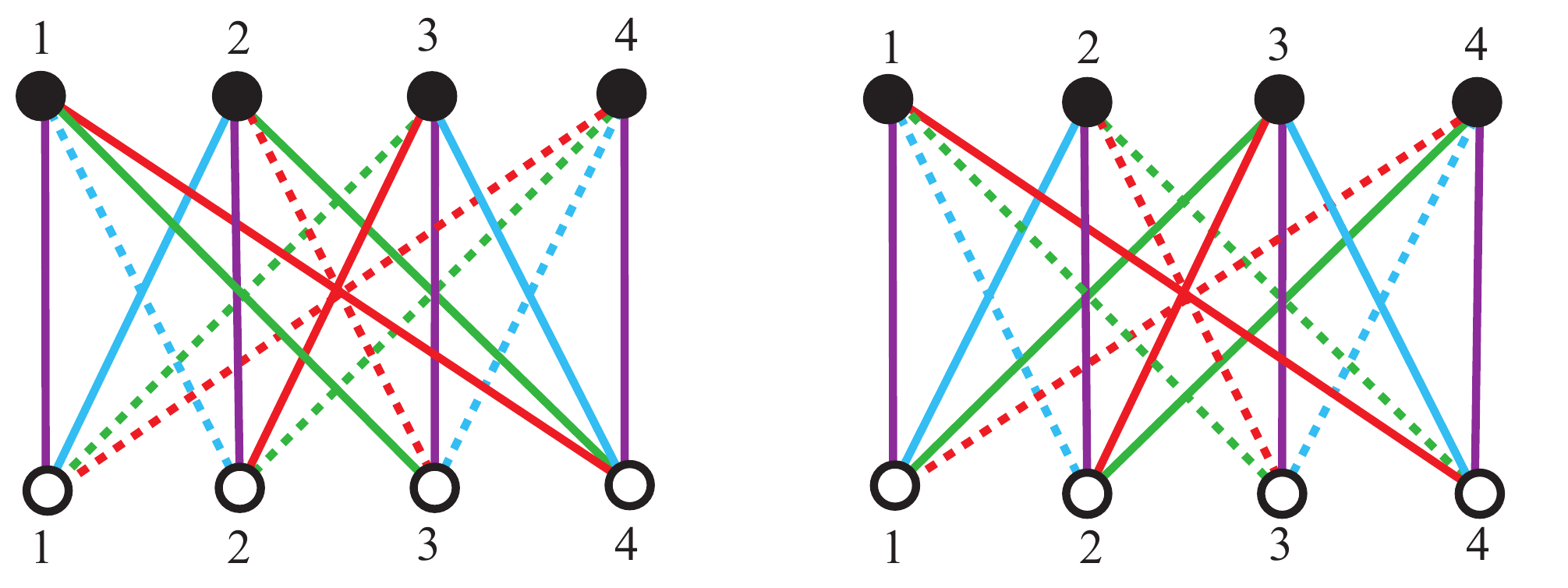}
 \caption{The two $N=4$ Valise Adinkras. The Adinkra on the right is obtained from the Adinkra on the left
 by implementing a ``twist", toggling the parity of the green edges.}
 \label{val4}
 \end{center}
 \end{figure}

 The smallest Adinkras which can possibly enhance to describe 4D
 supersymmetry are $N=4$ Adinkras describing 4+4 off-shell degrees
 of freedom.
 We therefore start by considering $N=4$ bosonic 4-4 Valise Adinkras.  There are exactly two of
 these not interrelated by cosmetic field redefinitions.
 These are exhibited in Figure \ref{val4}.
 In this paper we correlate the four edge colors with choices of the
 index $A$ so that {\purple purple}, {\color{blue} blue}, {\color{green} green}, and
{\color{red} red}
 correspond respectively to the operators $Q_{{\purple 1},{\blue 2},{\green 3},{\red
 4}}$.  For purposes of setting a convention for ordering the rows and columns of
 our linkage matrices, we sequence the boson fields $\phi_i$ and the
 fermion fields $\psi_{\hat{\imath}}$ using the obvious faithful correspondence
 with the index choices.  Furthermore, in the Adinkras exhibited in this section,
 the white vertices labeled 1, 2, 3, 4 represent the boson fields $\phi_i$
 with corresponding index choices, while
 the black vertices labeled 1, 2, 3, 4 represent the
 fermion fields $\psi_{\hat{\imath}}$ with corresponding index
 choices.  This allows us to readily translate
 each Adinkra into precise linkage matrices, using the technology
 explained in Appendix \ref{adinkrastuff}.

 The linkage matrices $(\,u_A\,)_i\,^{\hat{\imath}}$ corresponding to the first Adinkra in Figure
 \ref{val4} are exhibited in Table \ref{links44}.\footnote{The diligent reader
 should verify the correspondence between Figure \ref{val4} and Table
 \ref{links44} using the simple technology explained in Appendix \ref{adinkrastuff}.}
 These codify the ``upward" links connecting the bosons $\phi_i$ to the fermions
 $\psi_{\hat{\imath}}$ having greater engineering dimension.  Since there are no edges
 linking downward from any of the boson vertices, it follows that
 $(\,d_A\,)_i\,^{\hat{\imath}}=0$ in this case.  Similarly,
 we have $(\,\tilde{u}_A\,)_{\hat{\imath}}\,^i=0$, reflecting the fact that none
 of the fermions have upward directed edges.  Finally, we have
 $(\,\tilde{d}^A\,)_{\hat{\imath}}\,^j=\delta^{AB}\,\delta^{jk}\,(\,u_B\,)_k\,^{\hat{k}}\,\delta_{\hat{k}\hat{\imath}}$
 and
 $(\,d^A\,)_i\,^{\hat{\jmath}}=\delta^{AB}\,\delta^{\hat{\jmath}\hat{k}}\,(\,\tilde{u}_B\,)_{\hat{k}}\,^k\,\delta_{ki}$,
 schematically $\tilde{d}_A=u_A^T$ and $d_A=\tilde{u}_A^T$, reflecting the fact that every edge
 describes a pairing of an upward directed term and a corresponding downward
 directed term.
 The relationships $\tilde{d}_A=u_A^T$ and $d_A=\tilde{u}_A^T$ are characteristic of
 ``standard Adinkras".  Non-standard Adinkras, which can include ``one-way" upward
 Adinkra edges, appear in gauge multiplet shadows, as explained below, and
 in also in other contexts of interest.\footnote{Some considerations
 involving ``one-way" Adinkra edges were described
 in both \cite{RETM} and \cite{Frames}.}

 \subsection{The $N=4$ bosonic 4-4 Adinkras}
 Using the features $\tilde{u}_A=d^B=0$ and
 $\tilde{d}^A=u_A^T$, and using the matrices $G^a_{AB}$ in (\ref{Gmats}),
 we can begin to analyze the enhancement equations associated with the left Adinkra in Figure \ref{val4}.
 Consider the first equation in (\ref{gregeq}) for the index choices
 $(a|A,B)=(a|1,1)$.  Since $\Lambda^1_{11}=-G^1_{11}=0$, that
 $4\times 4$ matrix equation reads\, $u_1\,\tilde{\Delta}^1_1=0$.  We then use
 (\ref{ga}), along with the Gamma matrices in (\ref{gammats}) to
 determine $\tilde{\Delta}^1_1=-\tilde{\Delta}^0_3$, which is equivalent to
 $\tilde{\Delta}^1_1=-\tilde{d}_3$ using the nomenclature $\tilde{\Delta}^0_A\equiv \tilde{d}_A$.
 Thus, the first equation in (\ref{gregeq}) reduces for the left Adinkra in Figure 1 and
 the index selections $(a|A,B)=(1|1,1)$ to
 the simple matrix equation $u_1\,u_3^T=0$, where we have also used $\tilde{d}_3=u_3^T$.
 Using Table \ref{links44},
 it is easy to check that this simple requirement is not satisfied.
 This tells us that the left Adinkra in Figure \ref{val4} cannot enhance to
 a 4D non-gauge matter multiplet.  Since the linkage matrices
 associated with the right Adinkra in Figure \ref{val4} are obtained
 from Table \ref{links44} by toggling the overall sign on $u_3$
 only, and since the enhancement equation $u_1\,u_3^T=0$ is unchanged by such an operation,
 it follows that neither Adinkra in Figure \ref{val4} can
 enhance to a 4D non-gauge matter multiplet.

 \begin{table}
 \begin{center}
 \begin{tabular}{cc}
 $u_{\purple 1} \,\,=\,\, \ba{cc|cc}1&&&\\&1&&\\\hline &&1& \\&&&1\ea$ &
 $u_{\blue 2} \,\,=\,\, \ba{cc|cc} &1&& \\ -1&&&\\\hline &&&-1\\&&1&\ea$ \\
 $u_{\green 3} \,\,=\,\, \ba{cc|cc}&&-1&\\&&&-1\\\hline 1&&&\\&1&&\ea$ &
 $u_{\red 4} \,\,=\,\, \ba{cc|cc} &&&-1\\&&1& \\\hline &-1&&\\1&&&\ea$
 \end{tabular}
 \caption{The boson ``up" linkage matrices for the 4-4 Valise Adinkra shown in Figure
 \ref{val4}.}
 \label{links44}
 \end{center}
 \end{table}

 The methodology explained in the last paragraph can be applied
 systematically for each possible index choice $(a|A,B)$ for any
 selected Adinkra.  In each case the time-like linkage matrices
 $\Delta^a_{AB}$ are determined using (\ref{ga}), so that the
 enhancement equation can be translated to a matrix statement
 involving the linkage matrices specific to the 1D multiplet
 directly corresponding to the Adinkra.  In the following discussion
 we do not repeat most of these steps.  But the reader should be
 aware that equation (\ref{ga}) is used in each example which we
 discuss, and the use of this equation is what allows us to cast the
 enhancement equation in terms of the matrices $u_A$, $d_A=\Delta^0_A$, and
 their transposes.

 \subsection{The $N=4$ bosonic 3-4-1 Adinkras}
\begin{figure}
 \begin{center}
 \includegraphics[width=5in]{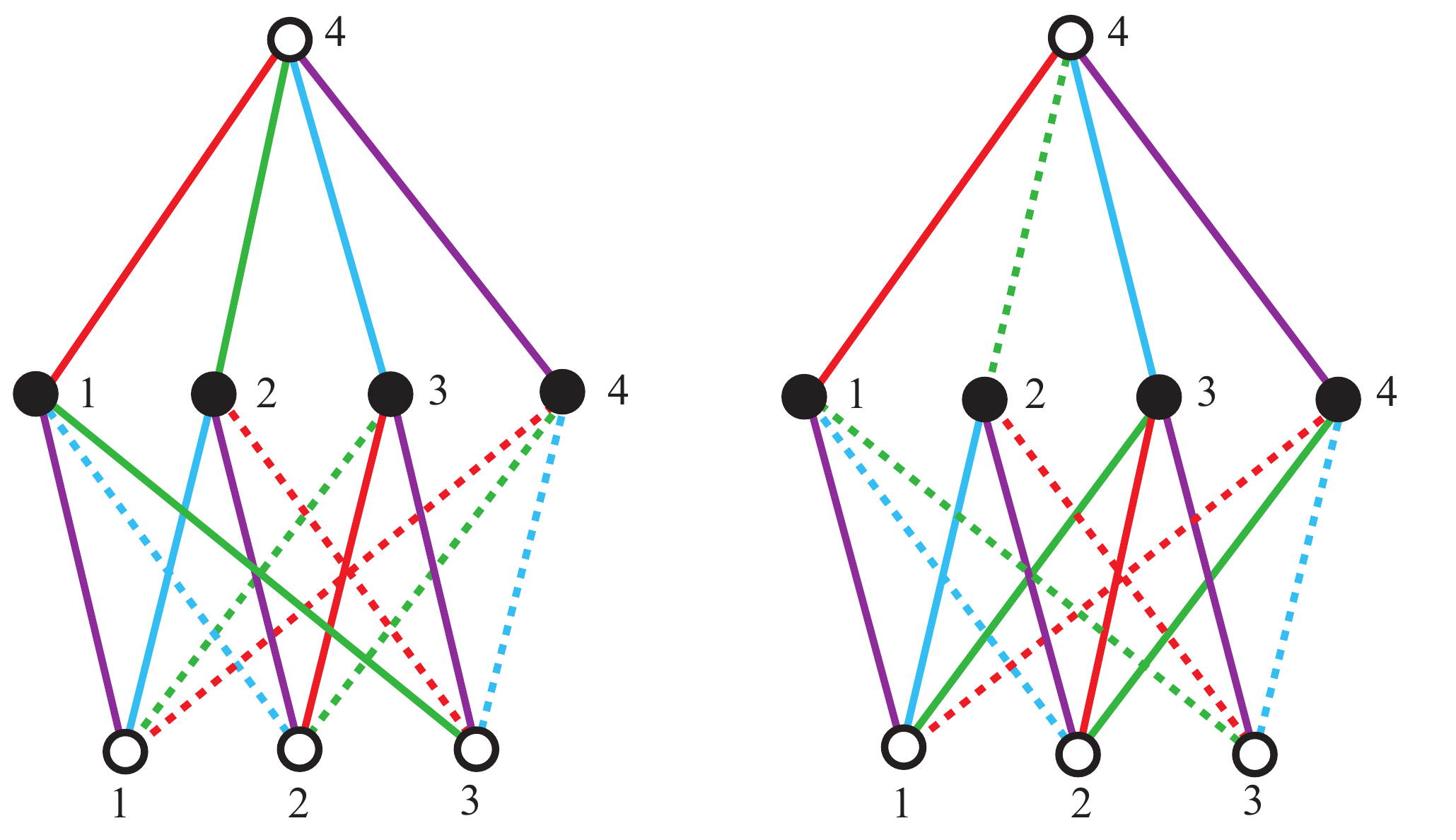}
 \caption{The two 3-4-1 Adinkras obtained from the Valise Adinkras in Figure \ref{val4}
 by raising one vertex. Here we have raised the boson vertex labeled 4.}
 \label{a341a}
 \end{center}
 \end{figure}

 Consider next those Adinkras obtained by raising one vertex
 starting with each Adinkra in Figure \ref{val4}.
 There are four possibilities starting from each of the two Valises,
 namely one possibility associated with raising any one of the
 four bosons.  For example, if we raise the boson vertex labeled
 ``4" starting from each Valise, what results are the two Adinkras in Figure \ref{a341a}.
 In these cases, we end up with three bosons at the lowest level, four
 fermions at the next level, and a single boson at the next level.
 We refer to Adinkras with this distribution of vertex
 multiplicities as bosonic 3-4-1 Adinkras, where the sequence of numerals
 faithfully enumerates the sequence of vertex multiplicities at
 successively higher levels.  (These alternate between boson and fermion
 multiplicities, naturally.)
 It is easy to see that there are exactly eight bosonic 3-4-1 $N=4$ Adinkras, and that these split evenly into
 two groups interrelated by a twist operation.

 We should point out that two Adinkras are equivalent if
 they are mapped into each other by cosmetic renaming of vertices,
 equivalent to linear automorphisms on the vector spaces spanned
 by the bosons $\phi_i$ or fermions $\psi_{\hat{\imath}}$, in cases where
 these maps preserve all vertex height assignments.  Such transformations have
 been called ``inner automorphisms". The simplest examples correspond to re-scaling any component field by a factor of $-1$.
 This manifests on an Adinkra by simultaneously toggling the parity
 of every edge connected to the vertex representing that field, \ie, by changing dashed edges into
 solid edges and vice-versa.  (This is referred to as ``flipping the vertex",
 and was described already in \cite{FG1}.) Our observation that there are two
 distinct families of minimal $N=4$ Adinkras interrelated by a twist
 operation refers to the readily-verifiable fact that one cannot ``undo" a twist by any inner
 automorphism.  (The curious reader might find it amusing to draw
 Adinkra diagrams, and investigate this statement for his or her
 self.)  It is also true that there are only two twist classes
 of minimal $N=4$ Adinkras, despite the fact that there are four different colors which can be
 used to implement a twist.  This is so because a given twist applied using any chosen edge
 color can be equivalently implemented as a twist applied using any other edge color augmented
 by a suitable inner automorphism.

  \begin{table}
 \begin{center}
 \begin{tabular}{cc}
 $u_{\purple 1} \,\,=\,\, \ba{cc|cc}1&&&\\&1&&\\\hline &&1&\\&&&0\ea$ &
 $d_{\purple 1} \,\,=\,\, \ba{cc|cc}0&&&\\&0&&\\\hline &&0& \\&&&1 \ea$ \\
 $u_{\blue 2} \,\,=\,\, \ba{cc|cc}&1&&\\-1&&&\\\hline &&&-1\\&&0& \ea$ &
 $d_{\blue 2} \,\,=\,\, \ba{cc|cc} &0&&\\0&&&\\\hline &&&0\\&&1& \ea$ \\
 $u_{\green 3} \,\,=\,\, \ba{cc|cc} &&-1& \\&&&-1\\\hline 1&&&\\&0&& \ea$ &
 $d_{\green 3} \,\,=\,\, \ba{cc|cc} &&0& \\&&&0\\\hline 0&&&\\&1&& \ea$ \\
 $u_{\red 4} \,\,=\,\, \ba{cc|cc}&&&-1\\&&1& \\\hline &-1&&\\0&&& \ea$ &
 $d_{\red 4} \,\,=\,\, \ba{cc|cc} &&&0\\&&0& \\\hline &0&&\\1&&& \ea$
 \end{tabular}
 \caption{Linkage matrices for the left 3-4-1 Adinkra shown in Figure
 \ref{a341a}.  The linkage matrices for the right Adinkra in that Figure are
 obtained from these by changing the sign of $u_{\green 3}$ and $d_{\green 3}$.}
 \label{links341}
 \end{center}
 \end{table}

 When we raise an Adinkra vertex, the up and down linkage matrices
 accordingly modify.   For example, consider the the 3-4-1 Adinkra on the
 left in Figure \ref{a341a},  obtained from the Adinkra on
 the left in Figure \ref{val4} by raising the $\phi_4$ vertex.
 The corresponding boson up and down matrices, which are straightforward to read
 off of the Adinkra, are shown in Table \ref{links341}.  Note that in this case the boson
 down matrices $d_A$ no longer vanish as they did in the case of the
 Valise.  This is because the $\phi_4$ vertex obtains downward links
 after being raised.  The fermion up matrices, which are determined for this standard
 Adinkra using $\tilde{u}_A=d_A^T$, are also non-vanishing after
 this vertex raise, since each of the fermions obtains an upward link to the
 boson $\phi_4$.

 Given a standard Adinkra, it is possible to raise the $n$-th boson
 vertex if and only if the $n$-th row of each boson down matrix
 is null; \ie,, provided $(\,d_A\,)_n\,^{\hat{\imath}}=0$
 for all values of ${\hat{\imath}}$.  This criterion ensures that
 the $n$-th boson vertex does not have any downward links which
 would preclude the vertex from being raised.
 (Since for standard Adinkras we have $\tilde{u}_A=d_A^T$, this criterion also
 implies that there are no lower fermions which link upward to the boson in question.)
 Absent such a tethering, the boson is free to be raised.  This operation is implemented
 algebraically by interchanging the
 $n$-th row of each boson up matrix $u_A$ with the $n$-th row of the
 respective boson down matrix $d_A$.  Thus, we implement the
 matrix reorganizations
 $(\,u_A\,)_n\,^{\hat{\imath}}\leftrightarrow
 (\,d_A\,)_n\,^{\hat{\imath}}$.  At the same time, we must
 interchange the $n$-th column of each fermion up matrix $\tilde{u}_A$ with the
 $n$-th column of the respective fermionic down matrix
 $\tilde{d}_A$, via
 $(\,\tilde{u}_A\,)_{\hat{\imath}}\,^n\leftrightarrow (\,\tilde{d}_A\,)_{\hat{\imath}}\,^n$.  The latter
 operation preserves the standard relationships $\tilde{u}_A=d_A^T$ and
 $\tilde{d}_A=u_A^T$.  It is easy to check that the linkage matrices
 in Table \ref{links341} are obtained from the linkage matrices in
 Table \ref{links44} by appropriately interchanging the fourth rows of the boson up and
 down matrices according to the above discussion.

 We now use the enhancement equation to analyze the eight standard $N=4$ bosonic 3-4-1
 Adinkras to ascertain if any of these can enhance to a 4D $N=1$ non-gauge matter multiplet.
 To begin, we start with the left Adinkra in Figure \ref{a341a}, by
 using the boson linkage matrices in Table \ref{links341} and the
 fermion linkage matrices determined by $\tilde{u}_A=d_A^T$ and $\tilde{d}_A=u_A^T$.  Using
 the $G^a_{AB}$ given in (\ref{Gmats}),
 the first condition in (\ref{gregeq}) reduces for the choice $(a|AB)=(1|11)$ to
 the matrix equation $u_1\,u_3^T+d_3\,d_1^T=0$.  Using the explicit matrices in Table \ref{links341},
 it is easy to see that this requirement is not satisfied.
 This tells us that the left Adinkra in Figure \ref{a341a} cannot enhance to
 a 4D non-gauge matter multiplet.

 \begin{figure}
 \begin{center}
 \includegraphics[width=5in]{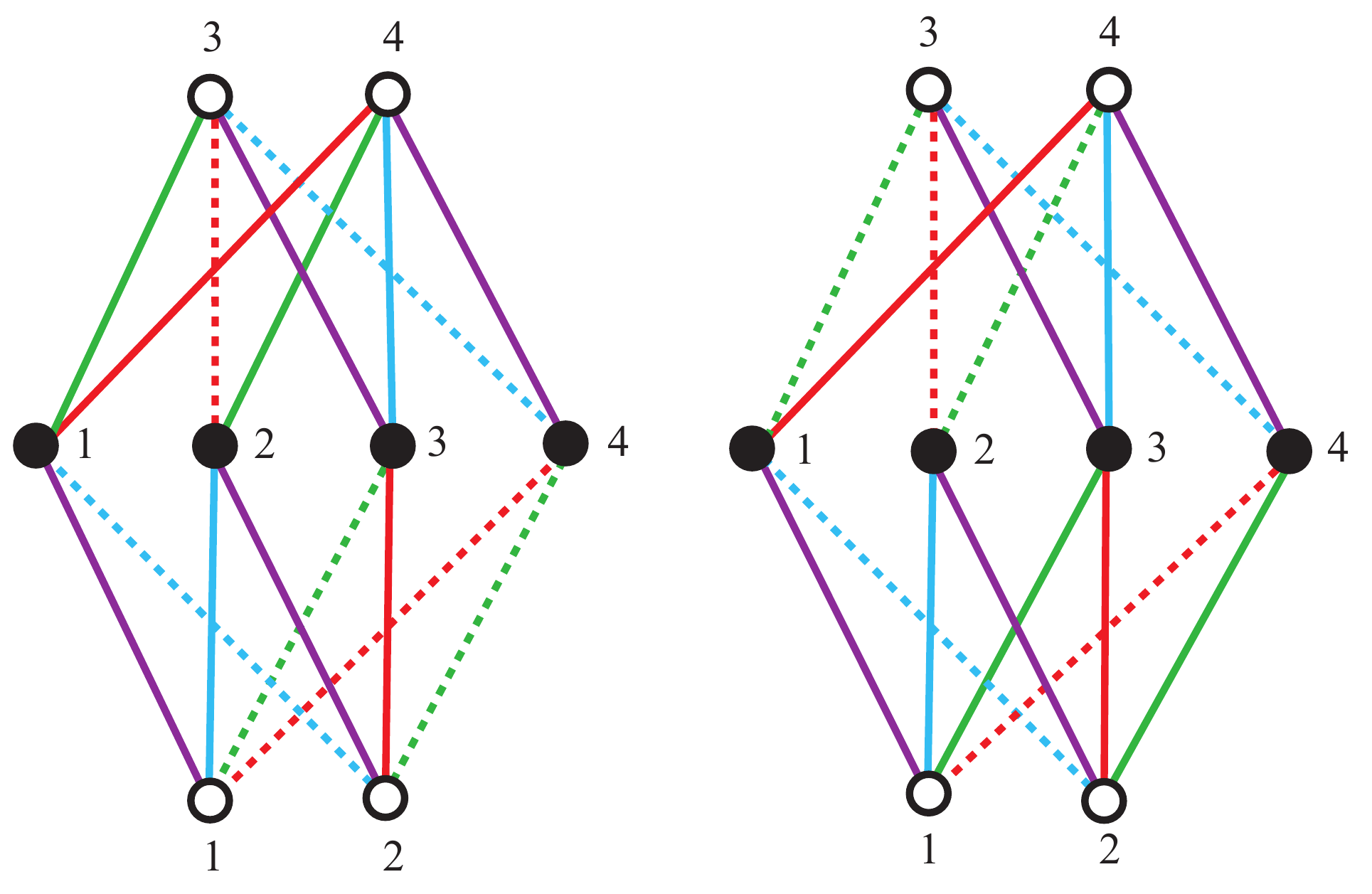}
 \caption{The $N=4$ 2-4-2 Adinkras may be obtained from 3-4-1 Adinkras by raising one
 vertex.  Here we have raised the third boson vertex starting with the two Adinkras
 shown in Figure \ref{a341a}.}
 \label{a242a}
 \end{center}
 \end{figure}

 Since the right Adinkra in Figure \ref{a341a} is obtained from the left Adinkra in that figure by
 twisting the green edges, corresponding to replacing $Q_3\to -Q_3$, the
 linkage matrices for that second 3-4-1 Adinkra are obtained from those in Table
 \ref{links341} by scaling the matrices $u_3$, $d_3$, $\tilde{u}_3$
 and $\tilde{d}_3$ each by a multiplicative minus sign.  The
 $(a|AB)=(1|1,1)$ enhancement equation, $u_1\,u_3^T+d_3\,d_1^T=0$,
 is unchanged by this operation.  So we conclude that neither Adinkra in
 Figure \ref{a341a} can enhance to a non-gauge 4D matter multiplet.
 It is straightforward to repeat this analysis for all cases
 associated with raising any possible single boson vertex starting
 with either of the Valise Adinkras in Figure \ref{val4}.  It
 follows, after careful analysis of each case, that the non-gauge enhancement
 equation (\ref{gregeq}) is not satisfied for any of the eight bosonic 3-4-1 Adinkras.

 \subsection{The $N=4$ bosonic 2-4-2 Adinkras}
 Things become more interesting when we raise one of the lower bosons in 3-4-1 Adinkras to obtain
 2-4-2 Adinkras.  In the end there are twelve minimal $N=4$ bosonic 2-4-2 Adinkras
 --- six obtained by two vertex raises starting from the
 left Adinkra in Figure \ref{val4} and six obtainable by two vertex raises starting from
 the right Adinkra in Figure \ref{val4}.  The six possibilities in each
 class correspond to the six different ways to select pairs from four
 choices.  For example, if we raise $\phi_3$ and $\phi_4$ in either case then what
 results are the two 2-4-2 Adinkras shown in Figure \ref{a242a}.
  For the left Adinkra in Figure \ref{a242a}, the boson linkage
 matrices are shown in Table \ref{links242}.  (It is straightforward to read
 these matrices off of the
 Adinkra. It is also straightforward to obtain these matrices algebraically, as explained above,
 by interchanging the third rows of the 3-4-1 up and down matrices shown in Table
 \ref{links341}.)

 \begin{table}
 \begin{center}
 \begin{tabular}{cc}
 $u_1 \,\,=\,\, \ba{cc|cc}1&&&\\&1&&\\\hline &&0&\\&&&0\ea$ &
 $d_1 \,\,=\,\, \ba{cc|cc}0&&&\\&0&&\\\hline &&1& \\&&&1 \ea$ \\
 $u_2 \,\,=\,\, \ba{cc|cc}&1&&\\-1&&&\\\hline &&&0\\&&0& \ea$ &
 $d_2 \,\,=\,\, \ba{cc|cc} &0&&\\0&&&\\\hline &&&-1\\&&1& \ea$ \\
 $u_3 \,\,=\,\, \ba{cc|cc} &&-1& \\&&&-1\\\hline 0&&&\\&0&& \ea$ &
 $d_3 \,\,=\,\, \ba{cc|cc} &&0& \\&&&0\\\hline 1&&&\\&1&& \ea$ \\
 $u_4 \,\,=\,\, \ba{cc|cc}&&&-1\\&&1& \\\hline &0&&\\0&&& \ea$ &
 $d_4 \,\,=\,\, \ba{cc|cc} &&&0\\&&0& \\\hline &-1&&\\1&&& \ea$
 \end{tabular}
 \caption{Linkage matrices for the left 2-4-2 Adinkra shown in Figure
 \ref{a242a}.  The linkage matrices for the right Adinkra in that Figure are
 obtained from these by changing the sign of $u_{\green 3}$ and $d_{\green 3}$.}
 \label{links242}
 \end{center}
 \end{table}

 We now use the enhancement equation (\ref{gregeq}) to analyze the twelve standard $N=4$
 2-4-2 Adinkras to ascertain if any of these can enhance to a 4D $N=1$ non-gauge matter
 multiplet. To begin, we start with the left Adinkra in Figure \ref{a242a}, equivalently
 described by the boson linkage matrices specified in Table
 \ref{links242} and by the fermion linkage matrices determined from these by
$\tilde{u}_A=d_A^T$ and $\tilde{d}_A=u_A^T$.

 We found above that for each of the two $N=4$ Valise Adinkras
 and for each of the eight 3-4-1 Adinkras the enhancement equation
 corresponding to $(a|A,B)=(1|1,1)$ is not satisfied.  This is
 equivalent to the statement that the $4\times 4$ matrix
 determined by $u_1\,u_3^T+d_3\,d_1^T$ does not vanish in these cases.  The reader
 should compute this combination in those cases, and then also
 compute this combination using the linkage matrices in Table
 \ref{links242}.  It is noteworthy that in this latter case, \ie, using the matrices in
 Table \ref{links242}, the computation of $u_1\,u_3^T+d_3\,d_1^T$
 does indeed produce the $4\times 4$ null matrix.  So the particular
 obstruction which we identified in the 4-4 and 3-4-1 Adinkras
 is notably absent for the specific 2-4-2 Adinkra shown on the left in
 Figure 1.

 Having satisfied the $(a|A,B)=(1|1,1)$ equation, it remains to analyze all of the
 other possible choices for $(a|A,B)$ and check the enhancement equations (\ref{gregeq})
 in each case.  It is interesting to comment on the
 case $(a|A,B)=(1|1,4)$, for example.  In this case the enhancement equation
 reads
 \brr u_1\,u_2^T+d_2\,d_1^T
      +u_4\,u_3^T+d_3\,d_4^T &=& 0 \,.
 \label{sho}\err
 Note that this is satisfied using the matrices in Table
 \ref{links242}.  So the left 2-4-2 Adinkra in Figure \ref{a242a}
 passes this second enhancement test.  (Thus, this Adinkra passes two out of
 60 different tests, counting the
 both the bosonic and the fermionic
 enhancement conditions for each of the 30 index choices $(a|A,B)$.)

 It is interesting that, unlike the left Adinkra in Figure \ref{a242a}, the right Adinkra in
 Figure \ref{a242a} fails the test (\ref{sho}).  This can be seen by noting that (\ref{sho})
 is sensitive to the parity on any one of the four edge colors since
 the overall signs on the first two terms flip upon toggling the
 sign on $Q_1$ or $Q_2$ while the sign on the third and fourth terms
 flip upon toggling the sign on $Q_3$ or $Q_4$.  More specifically,
 the linkage matrices for the second Adinkra
 in Figure \ref{a242a} are obtained from Table \ref{links242} by toggling the sign on $Q_3$,
 which toggles the overall sign on the matrices $u_3$ and $d_3$.  If we substitute the
 linkage matrices for the right Adinkra in Figure \ref{a242a}, obtained in this way, into (\ref{sho}) we find that
 this equation is no longer satisfied.  Thus, we conclude that the
 right Adinkra in Figure \ref{a242a} cannot enhance to a 4D non-gauge matter multiplet.
 Again, the reader should check these assertions by doing a few
 simple matrix calculations.

 Further analysis of all of the remaining 58 enhancement conditions
 shows that the matrices in Table \ref{links242} pass every one of
 these tests.  This is a non-trivial accomplishment, which indicates
 that the left Adinkra in Figure \ref{a242a} does represent the
 shadow of a 4D $N=1$ non-gauge matter multiplet.  As a
 representative example, consider the bosonic enhancement condition
 for the choice $(a|A,B)=(2|3,2)$.  This equation reads
 \brr u_2\,u_2^T+u_3\,u_3^T+d_2\,d_2^T+d_3\,d_3^T &=& 2 \,,
 \label{c232}\err
 where the factor of 2 on the right hand side means twice the $4\times 4$ unit matrix.
 The reader should
 verify that equation (\ref{c232}) is the bosonic enhancement
 equation described by (\ref{gregeq}) in this case.  The reader should
 also verify that (\ref{c232}) is satisfied by
 the linkage matrices in Table \ref{links242}.

 The fact that the first Adinkra in Figure \ref{a242a}
 describes the shadow of the 4D $N=1$ Chiral multiplet is easy to
 check by performing a direct dimensional reduction of the Chiral
 multiplet.  This is done explicitly in Appendix \ref{Chishadow}.
 In that Appendix we derive the shadow Adinkra, shown in Figure \ref{chshad},
 by direct translation of the 4D Chiral multiplet transformation rules.
 The left Adinkra in Figure \ref{a242a} is obtained from the Adinkra
 in Figure (\ref{chshad}) by reorganizing fields according to
 the following four permutation operations: $\phi_1\leftrightarrow\phi_2$,
 $\phi_3\leftrightarrow \phi_4$, $\psi_1\leftrightarrow\psi_4$, and
 $\psi_2\leftrightarrow\psi_3$.  This describes a cosmetic inner
 automorphism, indicating that the two Adinkras are equivalent.

 The fact that the left Adinkra in Figure \ref{a242a} passes the
 enhancement criteria while the right Adinkra does not identifies
 the left Adinkra as the Chiral multiplet shadow and identifies the
 right Adinkra as the so-called Twisted Chiral multiplet.  We also
 find that the 2-4-2 Adinkra obtained by raising the vertices
 $\phi_1$ and $\phi_2$ starting from the left 4-4 Valise in Figure 1 also
 passes all of the enhancement criteria whereas the twisted analog of
 this does not.  Scrutiny of all twelve bosonic 2-4-2 Valises
 confirms that only those two cases in the non-twisted family,
 associated with the left Valise in Figure \ref{val4}, obtained by raising either the pair
 $(\,\phi_1\,,\,\phi_2\,)$ or the pair $(\,\phi_3\,,\,\phi_4\,)$ can enhance to
 non-gauge matter multiplets in 4D.

 \subsection{The two 30-member $N=4$ Adinkra families}
 It is straighforward to systematically check the enhancement
 conditions for all 30 Adinkras in each of the two families ---
 one family associated with each of the two Valises in Figure \ref{val4}.  In each
 case, the 30-member family consists of the bosonic 4-4 Adinkra (the
 Valise), four bosonic 3-4-1 Adinkras, six bosonic 2-4-2 Adinkras, four
 bosonic 1-4-3 Adinkras, and the Klein flip of each one of these 15
 representatives.  (The Klein flipped Adinkras are the fermionic 4-4
 Valise and the 14 fermionic Adinkras obtainable from this by
 various vertex raises.)

 In total there are exactly four out of the 60 minimal Adinkras
 which pass our non-gauge enhancement criteria (\ref{gregeq}).
 The first two are the bosonic 2-4-2 Chiral multiplet shadows obtained by
 raising either $\phi_1$ and $\phi_2$ or by raising $\phi_3$ and
 $\phi_4$ starting from the left Adinkra in Figure \ref{val4}.
 The other two Adinkras reside in the other (relatively twisted)
 family, and are obtained from the right Adinkra in Figure \ref{val4}
 by first raising all four bosonic vertices, then
 raising either the fermionic vertices $\psi_1$ and $\psi_2$ or by raising
 the fermionic vertices $\psi_3$ and $\psi_4$.  These operations produce fermionic 2-4-2 Adinkras
 corresponding to twisted Klein flips of the two enhanceable bosonic 2-4-2 Adinkras.
 Since in these cases there are at most two fermions at any given
 height assignment, it is clear that these cannot assemble as 4D
 spinors.  As we explained above, the Adinkras which pass the
 enhancement criteria come in pairs, one element of which passes
 the spin-statistics test and one which does not.  In this way we
 conclude that of the 60 minimal Adinkras specified above only the
 two bosonic 2-4-2 cases can describe shadows of non-gauge 4D matter
 multiplets.

 It might appear curious that the four bosonic 2-4-2 Adinkras obtained from
 the right Adinkra in Figure \ref{val4} by raising $(\,\phi_1\,,\,\phi_3\,)$ or
 $(\,\phi_1\,,\,\phi_4\,)$ or $(\,\phi_2\,,\,\phi_3\,)$ or
 $(\,\phi_2\,,\,\phi_4\,)$ do not pass the enhancement criteria
 whereas the two Adinkras obtained by raising
 $(\,\phi_1\,,\,\phi_2\,)$ or $(\,\phi_3\,,\,\phi_4\,)$ do pass
 this test.  The reason why certain combinations of component fields
 appear favored relates to the fact that we have made a
 choice of spin structure when we selected the particular Gamma matrices in
 (\ref{gammaj}).  It is interesting that we lose no generality in
 making such a choice, however, since the freedom to choose a 4D spin basis
 is replaced by a corresponding freedom to perform inner
 automorphisms on the vector space spanned by the 1D component
 fields.

 Upon selecting a higher-dimensional spin basis, the enhancement equations (\ref{gregeq}) place
 restrictions on the component fields which are legitimately
 meaningful; the result that exactly two out of the 60 minimal $N=4$ Adinkras enhance to
 non-gauge 4D supersymmetric matter, along with the observation that
 those 1D multiplets which do enhance have 2-4-2 component
 multiplicities says something salient about 4D supersymmetry
 representation theory.  Specifically it says that any 4D $N=1$
 non-gauge matter multiplet must have two physical bosons, four
 fermions, and two auxiliary bosons.  This corroborates
 what has long been known about the minimal representations of 4D $N=1$ supersymmetry.
 What is remarkable is that we have hereby shown that this
 information is fully extractable using merely 1D supersymmetry and a
 choice of 4D spin structure --- that this information lies fully
 encoded in the 1D supersymmetry representation theory codified by
 the families of Adinkras, and that the key to unlocking this
 information is contained in our enhancement equation
 (\ref{gregeq}).\footnote{We believe that the extra
 structures, namely that the bosons complexify and that the fermions
 assemble into chiral spinors, is also encoded in our formalism,
 using the equations (\ref{tcons}). We also believe that deeper scrutiny of those equations should provide
 an algebraic context for broadly resolving natural organizations of
 supermultiplets, including complex structures, quaternionic structures, and so forth,
 in diverse dimensions.  But this lies
 beyond the scope of this introductory paper on this topic.}

 \section{Gauge multiplets}
 \label{phan}
 The non-gauge enhancement condition (\ref{gregeq}) relies on the result
 (\ref{ga}), which is derived in Appendix A.
 An important part of that derivation uses the assumptions
 $\Delta^0_A=\tilde{u}^T$ and $\tilde{\Delta}^0_A=u_A^T$.  These
 translate into the statement that every Adinkra edge codifies both
 an upward-directed term and a downward-directed term in the
 multiplet transformation rules.  (In other words, this result
 applies to ``standard" Adinkras.)
 But the presence of gauge degrees of freedom or Bianchi identities
 obviates this assumption.  This is demonstrated explicitly by dimensionally reducing the
 4D $N=1$ Maxwell field-strength multiplet, as described in detail in Appendix \ref{maxlinks}.

 \subsection{Introducing Phantoms}
 In field strength multiplets,  the vector
 space spanned by the boson components $\phi_i$ is
 larger than the vector space spanned by the fermion components
  $\psi_{\hat{\imath}}$.  The physical degrees of freedom
 balance, however, owing to redundancies in the space of bosons,
 related to the constraints.  This feature manifests in
 non-square linkage matrices, including sectors which decouple on
 the shadow.  We call these ``phantom sectors".

 The Maxwell multiplet is characteristic of generic
 multiplets involving closed $p$-form field strengths, when $p\ge 2$.
 In these cases, the field strength divides into an ``electric" sector,
 including components with a time-like index, and a ``magnetic" sector involving
 components which have only space-like indices.  The electric sector
 and the magnetic sector are correlated by the differential
 constraints implied by the Bianchi identity.  Upon reduction to
 one-dimension, the magnetic sector decouples.  The reason
 for this is that locally the magnetic fields are pure space
 derivatives, which vanish upon restriction to a zero-brane.

 Thus, in order to enhance a one-dimensional gauge multiplet to a
 higher-dimensional analog, not only do we have to resurrect the
 spatial derivatives, $\der_a$, but we also have to resurrect the
 gauge sector.  In the case of field strength multiplets, this means
 re-instating the magnetic fields.  Since these are physically
 decoupled on the shadow, they are re-introduced in an interesting
 way, at the top of one-way upward-directed Adinkra edges.  These nascent
 magnetic vertices play no role in the one-dimensional supersymmetry algebra
 respected by the other fields on the shadow.  (But they play an important role
 in closing the algebra in ambient higher dimensions.)
 We therefore call these vertices ``phantoms".

 Phantoms may be introduced into one-dimensional supermultiplets to
 enable possible enhancement involving closed $p$-form multiplet components.  Since closed 1-forms
 do not involve gauge invariance, it follows that the simplest case involves closed
 2-forms, such as $F_{\mu\nu}$ subject to $\der_{[\lambda}F_{\mu\nu]}=0$.
 This allows access to the important cases involving Vector
 multiplets. The higher-$p$ cases may be treated similarly,
 but these involve additional subtlety.  In order to keep our
 presentation relatively concise, we will not address cases $p\ge 3$
 nor will we address cases involving gauge fermions.  Our discussion
 will remain focussed on the ability to include 4D Abelian field
 strengths.  We also avoid other
 subtle technicalities by allowing 4D fermions to assemble only as
 spin 1/2 fields; that is we will not address the case of spin 3/2,
 or Rarita-Schwinger fields, in this paper.  It is straightforward to
 generalize our technology to allow for these possibilities.  But we
 defer discussions of these cases to future work, for reasons of
 bounding complexity.

 The structure of a phantom sector is usefully codified by phantom link matrices, defined as
 \brr (\,P_A\,)_i\,^{\hat{\imath}} &:=&
      (\,\tilde{u}_A^T-\Delta^0_A\,)_i\,^{\hat{\imath}} \,,
 \label{phantasm}\err
 where $\tilde{u}_A^T$ is the transpose of the $A$-th fermion ``up" matrix.
 A nonvanishing phantom matrix indicates the presence of one-way
 upward-directed Adinkra edges.  If $P_A$ is non-vanishing then this
 modifies the analysis in Appendix A precisely at the point where
 (\ref{faa}) is introduced as the transpose of (\ref{util}).
 If the Phantom matrices are included and the analysis is repeated,
 it is easy to show that (\ref{ga}) generalizes to
 \brr \Delta^a_A &=& -(\,\Gamma^0\Gamma^a\,\Delta^0\,)_A
      -\fr12\,(\,\Gamma^0\Gamma^a\,P\,)_A
      +T^{0a}\,P_A-P_A\,\tilde{T}^{0a} \,.
 \label{gaugeboo}\err
 Note that the final three terms will contribute nontrivially to this equation
 only in the gauge sector.

 It is helpful to briefly review the particular phantom sector
 associated with the shadow of the Maxwell field strength multiplet.
 This provides the archetype for generalizations, and motivates
 what follows.

 \subsection{Maxwell's shadow}
 The 4D $N=1$ super Maxwell multiplet involves four boson degrees of
 freedom off-shell.
 These organize as the auxiliary scalar $D$ plus the three off-shell
 ``electromagnetic" degrees of freedom described by $F_{\mu\nu}$.
 It is natural to write $E_a=F_{0a}$ and
 $B^a=\fr12\,\ve^{abc}\,F_{bc}$.  The Bianchi identity
 $\der_{[\lambda}F_{\mu\nu]}=0$ correlates $E_a$ and $B^a$.
 Locally, we can solve the Bianchi identity in terms of a vector
 potential $A_\mu$, so that $E_a=\der_0A_a-\der_aA_0$ and
 $B^a=\ve^{abc}\,\der_bA_c$.  Upon restriction to the zero-brane we
 take $\der_a\to 0$, so that $E_a\to\der_0A_a$ and $B_a\to 0$.
 Since the magnetic fields vanish on the zero-brane, it is
 natural to think of the $E_a$ as more fundamental for our purposes.
 The shadow is described by a fermionic 4-4 Adinkra where the bosons
 are $(\,E_1\,,\,E_2\,,\,E_3\,,\,D\,)$ and the fermions are
 $(\,\lambda_1\,,\,\lambda_2\,,\,\lambda_3\,,\,\lambda_4\,)$.
 To enhance this multiplet we must re-introduce $\der_a$ and also
 re-introduce the fields $B^a$, along with constraints.  To do this, we
 allow for ``phantom" bosons on the worldline, which
 correspond to the $B^a$ off of the worldline.

 To accommodate the phantom bosons, we consider an enlarged bosonic
 vector space, $\phi_i=(\,E_1\,,\,E_2\,,\,E_3\,,\,D\,|\,B_1\,,\,B_2\,,\,B_3\,)$
 in conjunction with the fermionic vector space
 $\psi_{\hat{\imath}}=(\,\lambda_1\,,\,\lambda_2\,,\,\lambda_3\,,\,\lambda_4\,)$.
 As a useful index notation, we write these as
 $\phi_i=(\,E_a\,,\,D\,|\,B^{\bar{a}}\,)$, where $B^{\bar{a}}$ is the magnetic
 phantom associated with $E_a$.  Thus, $a$ and $\bar{a}$ each assume the values 1,2,3,
 and we have $\phi_{1,2,3}=E_{1,2,3}$ and $\phi_{5,6,7}=B^{\bar{1},\bar{2},\bar{3}}$, respectively.
 In this way, phantom fields are designated by an over-bar on the relevant
 index.  Boson fields not in the phantom sector are indicated by underlined indices, so that
 $\phi_{\underline{1},\underline{2},\underline{3},\underline{4}}=(\,E_1\,,\,E_2\,,\,E_3\,,\,D\,)$.
 Matrices with two boson indices then divide into four sectors,
 $X_{\underline{i}}\,^{\underline{j}}$,
 $X_{\underline{i}}\,^{\bar{a}}$, $X_{\bar{a}}\,^{\underline{j}}$,
 and $X_{\bar{a}}\,^{\bar{b}}$.

 The shadow transformation rules associated with the Maxwell
 multiplet can be written as (\ref{ssm}), but the linkage matrices
 are not square!  Instead,
 $(\,\tilde{u}_A\,)_{\hat{\imath}}\,^j$ is $7\times 4$ and
 $(\,\Delta^\mu_A\,)_i\,^{\hat{\imath}}$ is $4\times 7$.
 We exhibit the precise linkage matrices
 associated with the Maxwell multiplet in Appendix \ref{maxlinks}.
 For the Super Maxwell case, the first enhancement equation in (\ref{gregeq}) is
 a $7\times 7$ matrix equation, whereas the
 second is a $4\times 4$ matrix equation.
 The first equation has phantom sectors which can be shuffled away canonically
 via use of the Bianchi identity, as explained below.

 \subsection{Canonical Re-shuffling}
 Owing to the derivatives in the enhancement condition (\ref{gregeq}), we may use
 the Bianchi identity, $\der_{[\lambda}F_{\mu\nu]}=0$, usefully
 re-written as
 \brr \der_0\,B^{\bar{a}} &=& \ve^{\bar{a}bc}\,\der_b\,E_c
      \nonumber\\[.1in]
      \der_{\bar{a}}\,B^{\bar{a}} &=& 0 \,,
 \label{bz}\err
 to define ``canonical reorganizations" of the matrices in
 (\ref{matos}) under which (\ref{gregeq}) remains unchanged.
 Specifically, the first equation in (\ref{bz}) allows us to redefine
 \brr (\,\Omega^0_{AB}\,)_i\,^{\bar{a}} &\to& 0
      \nonumber\\[.1in]
      (\,\Omega^a_{AB}\,)_i\,^b &\to&
      (\,\Omega^a_{AB}\,)_i\,^b
      +\ve^{ab}\,_{\bar{c}}\,(\,\Omega^0_{AB}\,)_i\,^{\bar{c}} \,,
 \label{can1}\err
 whereby we exchange each appearance of $\der_0\,B^{\bar{a}}$ in a supersymmetry commutator
 with an equivalent
 expression involving spatial derivatives on the electric
 field components.
 Similarly, the second equation in (\ref{bz}) allows us to redefine
 \brr (\,\Omega^1_{AB}\,)_i\,^{\bar{1}} &\to& 0
      \nonumber\\[.1in]
      (\,\Omega^2_{AB}\,)_i\,^{\bar{2}} &\to&
      (\,\Omega^2_{AB}\,)_i\,^{\bar{2}}
      -(\,\Omega^1_{AB}\,)_i\,^{\bar{1}}
      \nonumber\\[.1in]
      (\,\Omega^3_{AB}\,)_i\,^{\bar{3}} &\to&
      (\,\Omega^3_{AB}\,)_i\,^{\bar{3}}
      -(\,\Omega^1_{AB}\,)_i\,^{\bar{1}} \,.
 \label{can2}\err
 In this way, we define a canonical structure of the matrices
 $(\,\Omega^\mu_{AB}\,)_i\,^j$, ensured by the transformations
 (\ref{can1}) and (\ref{can2}), enabled by the Bianchi identity
 (\ref{bz}), whereby $(\,\Omega^0_{AB}\,)_i\,^{\bar{a}}=0$ and
 $(\,\Omega^1_{AB}\,)_i\,^1=0$.  The first equation in (\ref{gregeq})
 may now be interpreted as saying that
 $(\,\Omega^\mu_{AB}\,)_i\,^j\to 0$ under these transformations.

 We remark that that each of the 40 $7\times 7$ matrices
 $(\,\Omega^\mu_{AB}\,)_i\,^j$ defined by (\ref{matos}), using the linkage
 matrices exhibited in Appendix \ref{maxlinks}, do satisfy
 $(\,\Omega^\mu_{AB}\,)_i\,^j\to 0$ using the transformations
 (\ref{can1}) and (\ref{can2}).  The diligent reader is
 encouraged to check this assertion.

 \subsection{The $p=1$ gauge enhancement conditions}
 Based on the above, a means becomes apparent under which we can ascertain which one-dimensional
 multiplets may enhance to 4D gauge field strength multiplets, based only on
 a knowledge of the one-dimensional transformation rules, or
 equivalently given an Adinkra.

 For physical gauge fields, the bosonic field strength tensor
 has greater engineering dimension than the corresponding gaugino fermions.
 Therefore, the ambient fermions transform into the magnetic fields via
 terms in the fermion transformation rule $\delta_Q\,\lambda$ given by
 $\fr12\,\ve_{abc}\,B^a\,\Gamma^{bc}\,\e$ or by
 $\fr12\,\ve_{abc}\,B^a\,\Gamma^{bc}\Gamma_5\,\e$.  These are the
 only Lorentz covariant possibilities.  The former case involves a
 vector potential and the latter case involves an axial vector
 potential.  We focus first on the former case, and comment on axial
 vectors afterwards.  It is straightforward to determine the phantom
 ``up" links and the time-like fermion ``down" links using $\delta_Q\,\lambda=\cdots
 +\fr12\,\ve_{abc}\,B^a\,\Gamma^{bc}\,\e$.  By rearranging this term
 into the form $\delta_Q\,\lambda_i=\cdots
 +\e^A\,(\,\tilde{u}_A\,)_i\,^{\bar{a}}\,B_{\bar{a}}$, we derive
\footnote{Consistency of (\ref{u0yea}) with (\ref{phantasm}) implies
 usefully extractable information about the spin representation
 assignments.  We will not explore this arena in this paper.}
 \brr (\,\tilde{u}_A\,)_{\hat{\imath}}\,^{\bar{a}} &=&
      \fr12\,\ve^{\bar{a}bc}\,(\,\Gamma_{bc}\,)_{\hat{i}A}
      \nonumber\\[.1in]
      (\,\Delta^0_A\,)_{\bar{a}}\,^{\hat{\imath}} &=& 0 \,,
 \label{u0yea}\err
 whereby using (\ref{phantasm}) we determine the non-vanishing part of the phantom matrix as
 \brr (\,P_A\,)_{\bar{a}}\,^{\hat{\imath}} &=&
       \fr12\,\ve_{\bar{a}bc}\,(\,\Gamma^{bc}\,)_A\,^{\hat{\imath}}
       \,.
 \label{maxwphan}\err
 The entire phantom matrix has non-vanishing entries only in its
 final three rows.

 We can resolve the $\Delta^a_A$ matrices in two parts, using different methods.
  First we
 resolve the phantom part
 $(\,\Delta^a_A\,)_{\bar{a}}\,^{\hat{\imath}}$. Then we resolve the
 non-phantom part $(\,\Delta^a_A\,)_{\underline{a}}\,^{\hat{\imath}}$.

 We determine the phantom part of the space-like boson down matrices
 using the fact that
 $(\,\Delta^0_A\,)_{\bar{a}}\,^{\hat{\imath}}=0$, which says that there
 are no connections linking downward from the phantoms.  Thus, equation
 (\ref{daa}) tells us $(\,\Delta^a_A\,)_{\bar{a}}\,^{\hat{\imath}}=
 -(\,T^{0a}\,)_{\bar{a}}\,^i\,(\,\Delta_A^0\,)_i\,^{\hat{\imath}}$.
 Next, we use the fact that a boost shuffles magnetic fields into electric
 fields, via
 $(\,T_{0a}\,)_b\,^{\bar{c}}\,B_{\bar{c}}=\ve_{ab}\,^c\,E_c$,
 to determine
 \brr (\,\Delta^a_A\,)_{\bar{b}}\,^{\hat{\imath}} &=&
      \ve_{\bar{b}}\,^{a\underline{c}}\,(\,\Delta^0_A\,)_{\underline{c}}\,^{\hat{\imath}}
      \,.
 \label{ddyea}\err
 This determines the $\bar{b}$-th row in the phantom sector of the
 $A$-th space-like down matrices in terms of ``electric" rows in the
 time-like down matrices.

 We determine the non-phantom part of the space-like down matrices
 by considering the non-phantom sector in (\ref{gaugeboo}).  Thus, we allow only non-phantom values for the
 suppressed boson index.  In this case, the
 second and the
 fourth terms on the right-hand side vanish because
 $(\,P_A\,)_{\underline{a}}\,^{\hat{\imath}}=0$. The third term on
 the right-hand side can be resolved by noting that a boost shuffles electric fields into magnetic fields, via
 $(\,T^{0a}\,)_b\,^c\,E_c=\ve_b\,^{a\bar{c}}\,B_{\bar{c}}$,
 whereby $(\,T^{0a}\,)_{\underline{b}}\,^c=\ve_{\underline{b}}\,^{a\bar{c}}$. Substituting this result, along
 with (\ref{maxwphan}), we derive
 \brr (\,\Delta^a_A\,)_{\underline{b}}\,^{\hat{\imath}} &=&
      -(\,\Gamma^0\Gamma^a\,)_A\,^B\,(\,\Delta^0_A\,)_{\underline{b}}\,^{\hat{\imath}}
      +(\,\Gamma^a\,_{\underline{b}}\,)_A\,^{\hat{\imath}} \,.
 \label{dd2yea}\err
 This is the same as our non-gauge result (\ref{ga}), modified by the
 second term.\footnote{Note that the second term in (\ref{dd2yea})
 can also be written as $-2\,\ve_{\underline{b}}\,^{ac}\,(\,{\cal
 R}_c\,)_A\,^{\hat{\imath}}$, where ${\cal R}_c$ generates a
 rotation in the $c$-th spatial dimension.}
 Taken together, (\ref{ddyea}) and (\ref{dd2yea}) generate for us
 the entire space-like phantom-modified down matrices, generalizing
 our earlier non-gauge result (\ref{ga}) to the case in which fields
 can assemble into closed 1-forms.

 Note that the Maxwell field strength shadow, including its
 phantom sectors, may be reproduced from the minimal $N=4$ Adinkras
 using the methods described in this section.  Presently, we explain
 a means to produce a representative in one of the two minimal $N=4$
 Adinkra families which demonstrably enhances to a 4D Maxwell multiplet.
 Importantly, we can verify
 that this Adinkra enhances to such a 4D multiplet using only one-dimensional reasoning.

 This is done by starting with
  right Adinkra in
 Figure \ref{val4}. (Note that this Adinkra is in the family twisted relative to the
 family which includes the Chiral multiplet shadow.)
  From this starting Adinkra, we raise all four boson vertices, to
 obtain a fermionic 4-4
 Adinkra with the four bosons at the higher level.  We then perform a
 permutation of the
 first and the fourth boson vertices, and a permutation of the the second and the
 fourth boson vertices.  We then flip both the first and the fourth boson
 vertices.  (To flip a vertex means to scale this by an overall minus sign.)
 Finally we flip the third fermion vertex.  We compute the time-like up matrices
 $\tilde{u}_A$ and the time-like down matrices $\Delta^0_A$ using the resultant
 fermionic 4-4 Adinkra.  We designate the first three boson vertices in this final
 orientation as our designated ``electric" components. We then apply (\ref{u0yea}), (\ref{ddyea}),
 and (\ref{dd2yea}) to append phantom sectors to these time-like linkage matrices, and to
 generate provisional space-like down links $\Delta^a_A$, including phantom sectors.
 What results from these operations are precisely the matrices shown
 in Appendix \ref{maxlinks}.  We next compute the $\Omega$ matrices using
 (\ref{matos}). Finally, we apply a canonical reshuffling of these
 $\Omega$-matrices, using (\ref{can1}) and (\ref{can2}).  Happily,
 we find that under this operation
 all of the matrices $(\,\Omega^\mu_{AB}\,)_i\,^j$ and
 $(\,\tilde{\Omega}^\mu_{AB}\,)_{\hat{\imath}}\,^{\hat{\jmath}}$
 vanish.   This illustrates that this representative set of
 operations produces an Adinkra which passes our non-trivial gauge enhancement test.

 If we repeat the above search allowing for axial vector potential,
 we would modify all equations in this subsection with an additional
 factor of $\Gamma_5$.  What we find is that every multiplet
 which enhances to provide a vector potential also enhances to
 provide an axial vector potential.  This is loosely similar to the
 situation involving Chiral versus Antichiral multiplets, which
 have identical shadows.  It follows that the both the Vector multiplet
 and the Axial Vector multiplet shadow lie in the family of Adinkras
 relatively twisted relative to the Chiral multiplet.  We will not
 exhibit separate equations for the Axial Vector case, leaving that
 as an exercise for the interested reader.

 \subsection{Algorithm}
 We have already explained that it is possible to
 systematically generate
 the linkage matrices for each member of the family associated with a given
 Valise. For each representative, we can sequentially select triplets of vertices as potential electric
 field components, and use equations (\ref{u0yea}), (\ref{ddyea}), and (\ref{dd2yea}) to
 generate postulate phantom sectors.  For each of these, we
 compute the relevant $\Omega$-matrices using (\ref{matos}), then shuffle these into
 canonical form using (\ref{can1}) and (\ref{can2}).  By selecting those Adinkras
 for which all the canonical $\Omega$-matrices obtained in this way vanish, we
 thereby obtain an algorithmic search for all multiplets in which vertices can assemble
 into closed 1-forms.  This search is guaranteed to locate those multiplets which do
 properly enhance. (N.B. we have
 explained in the previous paragraph an example which we know
 works.)

 In the case of closed 1-form multiplets, an enhanceable Adinkra
 exhibits a synergy between the postulated electric
 field components and the designated magnetic phantom sector, vis-a-vis
 the assignment of the component basis $\phi_i$.  This is because the enhancement criteria are
 sensitive to the basis choice on the component
 boson vector via the structure of our imposed phantom sector.\footnote{Note that
 designating the phantom sector using (\ref{ddyea}) and
(\ref{dd2yea})
 does not remove generality from the search, much as choosing 4D Gamma matrices
 does not remove generality, for reasons described above.  Instead,
 this removes redundancies from the answer set.} Practically, this requires, for a
 comprehensive algorithmic search for enhanceable Adinkras, that in
 addition to sifting through all possible vertex raises and all
 possible selections of vertex triplets, we also have to sift
 through vertex permutations and vertex flips. Thus, inner automorphisms would seem
 to enlarge the effective search family.  Regardless, our discussion does
 show  that the portion of 4D supersymmetry representation theory
 involving closed 1-form multiplets is accessible and
 understandable using only 1D supersymmetry.  We find this
 interesting.

 In summary, following is a method to test an Adinkra to see if it enhances to give a
 4D multiplet with a closed 1-form gauge field strength:

 \noindent
 1) Select three boson vertices with a common engineering dimension as the presumed electric components,
 and arrange the boson vector space so that $\phi_{1,2,3}$
 correspond to these.\\[.1in]
 2) Compute time-like linkage matrices $u_A$, $\tilde{u}_A$,
 $\Delta^0_A$ and $\tilde{\Delta}^0_A$
 from the Adinkra.\\[.1in]
 3) Augment the boson vector space by
 adding on a phantom sector consisting of three new fields
 $\phi_{\bar{1},\bar{2},\bar{3}}$ with the same engineering dimension as
 $\phi_{1,2,3}$.\\[.1in]
 4) Add phantom sectors to the
 up matrices $\tilde{u}_A$, by adding
 three extra columns, and add phantom sectors to the time-like down matrices
 $\Delta^0_A$ by
 adding three extra rows.\\[.1in]
 5) Populate the phantom sector of the up matrices using
 (\ref{u0yea}).\\[.1in]
 6) Generate space-like down matrices, including phantom sectors
 using (\ref{ddyea}) and (\ref{dd2yea}).\\[.1in]
 7) Use the complete set of phantom-augmented linkage matrices to determine the matrices
 $(\,\Omega^a_{AB}\,)_i\,^j$ and $(\,\tilde{\Omega}^a_{AB}\,)_{\hat{\imath}}\,^{\hat{\jmath}}$
 using (\ref{matos}).\\[.1in]
 8) Reshuffle the boson matrix $(\,\Omega^a_{AB}\,)_i\,^j$
 using the prescription (\ref{can1}) and (\ref{can2}), to obtain a new matrix, in
 canonical form,
 \brr (\,\Omega^a_{AB}\,)_i\,^j &\to&
      (\,\widehat{\Omega}^a_{AB}\,)_i\,^j \,.
 \err
 The presence of the hat indicates canonical form.\\[.1in]
 9) The $p=1$ gauged enhancement conditions now correspond to the
 original enhancement conditions (\ref{gregeq}) augmented by the addition of
 phantom sectors and a canonical reshuffle.   We conclude that
 a necessary requirement for an Adinkra to enhance to a $p=1$
 field-strength multiplet is
  \brr (\,\widehat{\Omega}^\mu_{AB}\,)_i\,^j &=& 0
      \nonumber\\[.1in]
      (\,\tilde{\Omega}^\mu_{AB}\,)_{\hat{\imath}}\,^{\hat{\jmath}}
       &=& 0 \,.
 \label{gaugegreg}\err
 This is our $p=1$ gauge enhancement condition.  The key difference
 as compared to the non-gauge case is that the linkage matrices are
 not square in the gauge case, owing to the presence of the phantom
 boson sectors.  Furthermore, we must implement the canonical
 reshuffling maneuver to generate the hatted $\widehat{\Omega}$
 matrices which describe the non-gauge enhancement condition.

 The way we have designed our formalism is tailored toward implementation
 via computer-searches.  This may require supercomputers for cases
 with higher $N$, which will involve large matrix computations.
 We hope to enlarge our algorithms so that
 sifting through one-dimensional multiplets is controlled by the
 relevant lists of doubly-even error-correcting codes which
 correspond to these, as explained in the introduction.  But this
 lies beyond the scope of the presentation in this paper.  We find
 it sufficiently noteworthy that such algorithms exist, even in
 principle.  Our hope is that this will shed light on unknown
 aspects of supersymmetry which have defied attack using previous conventions.

 \section{Conclusions}
 We have presented algebraic conditions which allow one to
 systematically locate those representations of one-dimensional
 supersymmetry which may enhance to higher dimensions.
 Equivalently, we have explained how to discern whether a
 given one-dimensional supermultiplet is a shadow of a
 higher-dimensional analog.  This allows the representation
 theory of supersymmetry in diverse dimensions to be divided into
 the simpler representation theory of one-dimensional supersymmetry
 augmented with separate questions pertaining to the possibility of
 enhancement into higher dimensions.

 We have shown through explicit examples how information pertaining
 to four-dimensional $N=1$ supersymmetry may be extracted using
 only information from one-dimensional supersymmetry.  We did this comprehensively for the
 case of 4D $N=1$ non-gauge matter multiplets.  We have
 also explained how this systematics generalizes to cases involving
 higher-dimensional gauge invariances, specializing our discussion
 to the case of 4D $N=1$ Super-Maxwell theory.

 We intend to use the formalism and the algorithms developed above
 to seek inroads towards off-shell aspects of interesting
 supersymmetric contexts where the off-shell physics remains
 mysterious but potentially relevant.

 \appendix

 \renewcommand{\theequation}{A.\arabic{equation}}
 \section{A Proof}
 \label{prlemma}
 In this Appendix we prove that demanding Lorentz invariance of the
 linkage matrices defined in (\ref{ssm}) completely determines all
 of the space-like linkage matrices $\Delta^a_A$ in terms of the
 time-like linkage matrices $\Delta^0_A$, and does so in precisely
 the manner specified above as equation (\ref{ga}).  We also show
 how this same requirements provides constraints on the spin
 representation content of supermultiplet component fields.

 The linkage matrices $(\,\Delta^\mu_A\,)_i\,^{\hat{\imath}}$ transform under $\spin(1,D-1)$, manifestly, as
 \brr \delta_L\,(\,\Delta^\mu_A\,)_i\,^{\hat{\imath}} &=&
      \theta^\mu\,_\nu\,(\,\Delta^\nu_A\,)_i\,^{\hat{\imath}}
      +\fr14\,\theta^{\lambda\sigma}\,(\,\Gamma_{\lambda\sigma}\,)_A\,^B\,
      (\,\Delta^\mu_B\,)_i\,^{\hat{\imath}}
      \nonumber\\[.1in]
      & & +\fr12\,\theta^{\lambda\sigma}\,
      (\,T_{\lambda\sigma}\,)_i\,^j\,(\,\Delta^\mu_A\,)_j\,^{\hat{\imath}}
      -\fr12\,\theta^{\lambda\sigma}\,(\,\Delta^\mu_A\,)_i\,^{\hat{\jmath}}\,(\,\tilde{T}_{\lambda\sigma}\,)_{\hat{\jmath}}\,^{\hat{\imath}}
      \,.
 \label{tr}\err
 In (\ref{tr}), the first term indicates that on $(\,\Delta^\mu_A\,)_i\,^{\hat{\jmath}}$
 the $\mu$ index is a vector
 index, the second term indicates that the $A$ index is a spinor
 index, and the last line accommodates the representation content of
 the supermultiplet component fields.

 Using (\ref{tr}), we obtain the following boost
 and rotation transformations for the ``time-like" linkage matrices
 $(\,\Delta^0_A\,)_i\,^{\hat{\imath}}$,
 \brr \delta_{\rm boost}\,(\,\Delta^0_A\,)_i\,^{\hat{\imath}} &=&
      \theta^0\,_a\,(\,\Delta^a_A\,)_i\,^{\hat{\imath}}
      +\fr12\,\theta^{0a}\,(\,\Gamma_{0a}\,)_A\,^B\,
      (\,\Delta^0_B\,)_i\,^{\hat{\imath}}
      \nonumber\\[.1in]
      & & +\theta^{0a}\,
      (\,T_{0a}\,\Delta^0_A\,)_i\,^{\hat{\imath}}
      -\theta^{0a}\,(\,\Delta^0_A\,\tilde{T}_{0a}\,)_i\,^{\hat{\imath}}
      \nonumber\\[.1in]
      \delta_{\rm rotation}\,(\,\Delta^0_A\,)_i\,^{\hat{\imath}} &=&
      +\fr12\,\theta^{ab}\,(\,\Gamma_{ab}\,)_A\,^B\,(\,\Delta^0_B\,)_i\,^{\hat{\imath}}
      \nonumber\\[.1in]
      & &
      +\theta^{ab}\,(\,T_{ab}\,\Delta^0_A\,)_i\,^{\hat{\imath}}
      -\theta^{ab}\,(\,\Delta^0_A\,\tilde{T}_{ab}\,)_i\,^{\hat{\imath}}
      \,,
 \label{timetran}\err
 and the following boost and rotation transformations for the
 ``space-like" linkage matrices,
 \brr \delta_{\rm boost}\,(\,\Delta^a_A\,)_i\,^{\hat{\imath}} &=&
      \theta^a\,_0\,(\,\Delta^0_A\,)_i\,^{\hat{\imath}}
      +\fr12\,\theta^{0b}\,(\,\Gamma_0\Gamma_b\,)_A\,^B\,(\,\Delta^a_B\,)_i\,^{\hat{\imath}}
      \nonumber\\[.1in]
      & &
      +\theta^{0b}\,(\,T_{0b}\,\Delta^a_A\,)_i\,^{\hat{\imath}}
      -\theta^{0b}\,(\,\Delta^a_A\,\tilde{T}_{0b}\,)_i\,^{\hat{\imath}}
      \nonumber\\[.1in]
      \delta_{\rm rotation}\,(\,\Delta^a_A\,)_i\,^{\hat{\imath}} &=&
      \theta^a\,_b\,(\,\Delta^b_A\,)_i\,^{\hat{\imath}}
      +\fr14\,\theta^{bc}\,(\,\Gamma_{bc}\,)_A\,^B\,(\,\Delta^a_B\,)_i\,^{\hat{\imath}}
      \nonumber\\[.1in]
      & &
      +\fr12\,\theta^{bc}\,(\,T_{bc}\,\Delta^a_A\,)_i\,^{\hat{\imath}}
      -\fr12\,\theta^{bc}\,(\,\Delta^a_A\,\tilde{T}_{bc}\,)_i\,^{\hat{\imath}}
      \,.
 \label{spacetran}\err
 We demand that the linkage matrices are Lorentz invariant.
 This imposes that each of the transformations in (\ref{timetran}) and
 (\ref{spacetran}) must vanish.\footnote{This is
 similar to ``demanding" that the Gamma matrices appearing in a
 Salam-Strathdee superfield be Lorentz invariant --- in that case they
 are, automatically, as a consequence of the Clifford algebra.}
 Requiring $\delta_{\rm boost}\,\Delta^0=0$ imposes
  \brr (\,\Delta^a_A\,)_i\,^{\hat{\imath}} &=&
      -\fr12\,(\,\Gamma^0\Gamma^a\,)_A\,^B\,
      (\,\Delta^0_B\,)_i\,^{\hat{\imath}}
       -(\,T^{0a}\,\Delta^0_A\,)_i\,^{\hat{\imath}}
      +(\,\Delta^0_A\,\tilde{T}^{0a}\,)_i\,^{\hat{\imath}} \,.
 \label{daa}\err
 This determines $(\,\Delta^a_A\,)_i\,^{\hat{\imath}}$ in terms of
 $(\,\Delta^0_A\,)_i\,^{\hat{\imath}}$ and in terms of the
 representation assignments of the supermultiplet component fields.

 The remaining consequences of imposing Lorentz invariance on the
 linkage matrices $(\,\Delta^\mu_A\,)_i\,^{\hat{\imath}}$ are the
 following,
 \brr \fr12\,(\,\Gamma_{ab}\,)_A\,^B\,\Delta^0_B
      &=&
      \Delta^0_A\,\tilde{T}_{ab}
      -T_{ab}\,\Delta^0_A
      \nonumber\\[.1in]
      \delta_b\,^a\,\Delta^0_A &=&
      \fr12\,(\,\Gamma_0\Gamma_b\,)_A\,^B\,\Delta^a_B
      +T_{0b}\,\Delta^a_A
      -\Delta^a_A\,\tilde{T}_{0b}
      \nonumber\\[.1in]
      \eta^{a[b}\,\Delta^{c]}_A
      +\fr14\,(\,\Gamma^{bc}\,)_A\,^B\,\Delta^a_B
      &=& \fr12\,\Delta^a_A\,\tilde{T}^{bc}
      -\fr12\,T^{bc}\,\Delta^a_A  \,,
 \label{tcons}\err
 where the $(\,\cdot\,)_i\,^{\hat{\imath}}$ index structure has been
 suppressed on each term.
 These correspond, respectively, to $\delta_{\rm rotation}\,\Delta^0=0$,
 $\delta_{\rm boost}\,\Delta^a=0$, and $\delta_{\rm rotation}\,\Delta^a=0$, for arbitrary
 transformation parameters $\theta^{0a}$ and $\theta^{ab}$.  The
 equations (\ref{tcons}) place significant restrictions on the
 spin representation content of the component fields.
 As explained above, we suspect that these equations encode useful
 and extractable information regarding allowable complements of spin
 structures in supermultiplets in diverse dimensions.

 The linkage matrices $(\,u_A\,)_i\,^{\hat{\imath}}$ transform under
 $\spin(1,D-1)$, manifestly, as
 \brr \delta\,(\,u_A\,)_i\,^{\hat{\imath}} &=&
      \fr14\,\theta^{\mu\nu}\,(\,\Gamma_{\mu\nu}\,)_A\,^B\,(\,u_B\,)_i\,^{\hat{\imath}}
      +\fr12\,\theta^{\mu\nu}\,(\,T_{\mu\nu}\,u_A\,)_i\,^{\hat{\imath}}
      -\fr12\,\theta^{\mu\nu}\,(\,u_A\,\tilde{T}_{\mu\nu}\,)_i\,^{\hat{\imath}}
 \err
 Requiring that these transformations vanish imposes
 \brr \fr12\,(\,\Gamma_{\mu\nu}\,)_A\,^B\,(\,u_B\,)_i\,^{\hat{\imath}} &=&
      (\,u_A\,\tilde{T}_{\mu\nu}\,)_i\,^{\hat{\imath}}
      -(\,T_{\mu\nu}\,u_A\,)_i\,^{\hat{\imath}} \,.
 \err
 This indicates correlations between the up linkage matrices and the
 representation content of the component fields.

 Similar conditions result from
 demanding invariance of $(\,\tilde{u}_A\,)_{\hat{\imath}}\,^i$ and
 $(\,\tilde{\Delta}^\mu_A\,)_{\hat{\imath}}\,^i$.  These are
 obtained from the above constraints by placing tildes on all
 matrices which do not have tildes and removing tildes from those
 that do.
 For example, invariance of $(\,\tilde{u}_A\,)_{\hat{\imath}}\,^i$
 imposes
 \brr \fr12\,(\,\Gamma_{\mu\nu}\,)_A\,^B\,(\,\tilde{u}_B\,)_{\hat{\imath}}\,^i &=&
      (\,\tilde{u}_A\,T_{\mu\nu}\,)_{\hat{\imath}}\,^i
      -(\,\tilde{T}_{\mu\nu}\,\tilde{u}_A\,)_{\hat{\imath}}\,^i \,.
 \label{util}\err
 Note that for standard Adinkras we have $\Delta^0_A=\tilde{u}_A^T$.\footnote{
 For non-standard Adinkras, such as those which
 accommodate gauge invariances, this relationship does not hold.
 Instead, we define $\tilde{u}_A=\Delta^0_A+P_A$, where $P_A$ is a
 so-called Phantom matrix, which encodes the nexus of one-way
 upward-directed Adinkra edges.   This generalization is addressed
 in section \ref{phan}.}
 Thus, using (\ref{util}) we determine
 \brr \fr12\,(\,\Gamma_{\mu\nu}\,)_A\,^B\,(\,\Delta^0_B\,)_i\,^{\hat{\imath}} &=&
      (\,T^T_{\mu\nu}\,\Delta^0_A\,)_i\,^{\hat{\imath}}
      -(\,\Delta^0_A\,\tilde{T}^T_{\mu\nu}\,)_i\,^{\hat{\imath}} \,.
 \label{faa}\err
 This
 equation can be used in conjunction with (\ref{daa}) to
 replace that equation with an analog in which the representation
 matrices are not included.

 The boost matrices
 $(\,T_{0a}\,)_i\,^j$ and
 $(\,\tilde{T}_{0a}\,)_{\hat{\imath}}\,^{\hat{\jmath}}$
 are generically symmetric\footnote{For example, if the fermions assemble as spinors
 then $\tilde{T}_{0a}=\fr12\,\Gamma_0\Gamma_a$.  In the Majorana basis
 $\Gamma_0$ is
 antisymmetric and real while $\Gamma_a$ are symmetric and real, and since
 $\Gamma_0$ and $\Gamma_a$ anticommute, it follows that $\tilde{T}_{0a}$ is
 symmetric in that case. For vectors $V_a$ we have $(\,T_{0a}\,)_0\,^b=\delta_a\,^b$ and
 $(\,T_{0a}\,)_b\,^0=-\eta_{ab}$; for our metric choice $\eta_{ab}=-1$, so that
 these $T_{0a}$ are symmetric.  This reasoning generalizes
 to higher-rank tensors and to all products of tensor and spinor representations.
 (Note too, that if the boost matrices were antisymmetric, then (\ref{faa}) and (\ref{daa}) could be used
 together to prove the inconsistent result that $\Delta^a=0$.)}
 Therefore, (\ref{faa}) can be re-written as
 \brr (\,T_{0a}\,\Delta^0_A\,)_i\,^{\hat{\imath}}
      -(\,\Delta^0_A\,\tilde{T}_{0a}\,)_i\,^{\hat{\imath}} &=&
      \fr12\,(\,\Gamma_0\Gamma_a\,)_A\,^B\,(\,\Delta^0_B\,)_i\,^{\hat{\imath}}
      \,.
 \label{haa}\err
 Substituting this result for the last two terms in (\ref{daa}), we
 determine
 \brr (\,\Delta^a_A\,)_i\,^{\hat{\imath}} &=&
      -(\,\Gamma^0\Gamma^a\,)_A\,^B\,
      (\,\Delta^0_B\,)_i\,^{\hat{\imath}} \,.
 \label{wow}\err
 Remarkably, this relationship is completely independent of the
 representation content of the component fields.
 This is an interesting result, which says that the space-like
 linkage matrices $\Delta^a$ are determined from the time-like
 linkage matrices $\Delta^0$.

 \renewcommand{\theequation}{B.\arabic{equation}}
 \section{Adinkra Conventions}
 \label{adinkrastuff}
 In this Appendix we give a very concise overview of the
 graphical technology of Adinkra diagrams.
 These were introduced in
 \cite{FG1}, and have formed the basis of a multidisciplinary
 research endeavor, aimed at resolving a mathematically rigorous basis for supersymmetry
 \cite{DFGHIL01,AT1,Prepotentials,HDS,RETM,Frames}.
 Some of the conventions,
 notably as regards sign choices, have varied in these references,
 in part because some of these
 papers aim at a physics audience and some at a mathematics audience.
 Thus, one reason for this Appendix is to clarify our conventions, as used above,
 so that this paper can be appreciated without undue confusion.
 Another is to allow this paper to be functionally self-contained.

 A representation of $N$-extended supersymmetry in one time-like
 dimension consists of $d$ boson fields $\phi_i$ and
 $d$ fermion fields $\psi_{\hat{\imath}}$ endowed with a set
 of transformation rules, generated by $\delta_Q(\e)$, where $\e^A$ are a set of
 $N$ anticommuting parameters, which respect the $N$-extended
 supersymmetry algebra specified by the commutator
 $[\,\delta_Q(\e_1)\,,\,\delta_Q(\e_2)\,]=2\,i\,\delta_{AB}\,\e_1^A\,\e_2^B\,\der_\tau$.
 The transformation rules can be written for boson fields as
 $\delta_Q\,\phi_i=-i\,\e^A\,(\,Q_A\,)_i\,^{\hat{\imath}}\,\psi_{\hat{\imath}}$
 and for fermion fields as
 $\delta_Q\,\psi_{\hat{\imath}}=-i\,\e^A\,(\,\tilde{Q}_A\,)_{\hat{\imath}}\,^i\,\phi_i$,
 where $(\,Q_A\,)_i\,^{\hat{\imath}}$  and
 $(\,\tilde{Q}_A\,)_{\hat{\imath}}\,^i$ are two sets of $N$ abstract $d\times d$ matrix
 generators of supersymmetry.  By definition these represent
 \brr (\,Q_{(A}\,\tilde{Q}_{B)}\,)_i\,^j &=& i\,\delta_i\,^j\,\der_\tau
      \nonumber\\[.1in]
      (\,\tilde{Q}_{(A}\,Q_{B)}\,)_{\hat{\imath}}\,^{\hat{\jmath}} &=&
      i\,\delta_{\hat{\imath}}\,^{\hat{\jmath}}\,\der_\tau \,,
 \label{q1alg}\err
 where the symmetrization brackets are defined with
 ``weight-one".\footnote{
 whereby $X_{(A}\,Y_{B)}=\fr12\,(\,X_A\,Y_B+X_B\,Y_B\,)$.}

 It is possible to use cosmetic field
 redefinitions to re-define the component fields $\phi_i$ and
 $\psi_{\hat{\imath}}$ into a ``frame" where the generators
 $(\,Q_A\,)_i\,^{\hat{\jmath}}$ and
 $(\,\tilde{Q}_A\,)_{\hat{\imath}}\,^{\hat{\jmath}}$ are first order
 differential operators with a specialized matrix structure.  Specifically, it is
 possible to write
  \brr (\,Q_A\,)_i\,^{\hat{\imath}} &=&
      (\,u_A\,)_i\,^{\hat{\imath}}
      +(\,d_A\,)_i\,^{\hat{\imath}}\,\der_\tau
      \nonumber\\[.1in]
      (\,\tilde{Q}_A\,)_{\hat{\imath}}\,^i &=&
      i\,(\,\tilde{u}_A\,)_{\hat{\imath}}\,^i
      +i\,(\,\tilde{d}_A\,)_{\hat{\imath}}\,^i\,\der_\tau
      \,,
 \label{qqt}\err
 where $(\,u_A\,)_i\,^{\hat{\imath}}$, $(\,d_A\,)_i\,^{\hat{\imath}}$,
 $(\,\tilde{u}_A\,)_{\hat{\imath}}\,^j$, and
 $(\,\tilde{d}_A\,)_{\hat{\imath}}\,^j$ are four sets
 of $N$ real $d\times d$ ``linkage matrices" with the features that every entry
 of each of these matrices takes only one of three values, 0, 1, or
 $-1$, and such that there is at most one non-vanishing entry
 in every row and at-most one non-vanishing entry in every column of each of these matrices.  Remarkably, we
 lose no generality by specializing to generators of the sort
 (\ref{qqt}) with these particular properties.  A mathematical proof that any
 one-dimensional supermultiplet can be written in this manner is
 provided in \cite{DFGHIL01}.

 A simple example in the context of $N=2$ supersymmetry is given by
 the following transformation rules,
 \brr \delta_Q\,\phi_1 &=&
      -i\,\e^1\,\psi_1-i\,\e^2\,\psi_2
      \nonumber\\[.1in]
      \delta_Q\,\phi_2 &=&
      -i\,\e^1\,\der_\tau\psi_2
      +i\,\e^2\,\der_\tau\psi_1
      \nonumber\\[.1in]
      \delta_Q\,\psi_1 &=&
      \e^1\,\der_\tau\phi_1-\e^2\,\der_\tau\phi_2
      \nonumber\\[.1in]
      \delta_Q\,\psi_2 &=&
      \e^1\,\phi_1+\e^2\,\der_\tau\phi_2 \,.
 \label{ex2}\err
 It is straightforward to verify that these satisfy the commutator
 relationship specified above.

 The operator $\der_\tau$ carries unit engineering dimension,
 while supersymmetry parameters $\e^A$ carry engineering
 dimension one-half.\footnote{In a system where $\hbar=c=1$, a field
 with engineering dimension $q$ carries units of $(\,{\rm Mass}\,)^q$.}
 Thus, in order to balance units in the transformation rules
 (\ref{ex2}) it follows that the two fermions $\psi_{1,2}$ have a common engineering
 dimension one-half greater than $\phi_1$, and that $\phi_2$ has an
 engineering dimension one-half greater than the fermions, and one unit
 greater than $\phi_1$.

 The transformation rules (\ref{ex2}) can be expressed equivalently,
 in terms of linkage matrices, as
 \brr (\,u_1\,)_i\,^{\hat{\jmath}} &=&
      \ba{cc} 1& \\ & 0 \ea
      \hspace{.4in}
      (\,u_2\,)_i\,^{\hat{\imath}} \,\,=\,\,
      \ba{cc}&1\\0& \ea
      \nonumber\\[.1in]
      (\,d_1\,)_i\,^{\hat{\imath}} &=&
      \ba{cc}0&\\&1\ea
      \hspace{.3in}
      (\,d_2\,)_i\,^{\hat{\imath}} \,\,=\,\,
      \ba{cc}&0\\-1&\ea
      \nonumber\\[.1in]
      (\,\tilde{u}_1\,)_{\hat{\imath}}\,^j &=&
      \ba{cc}0&\\&1\ea
      \hspace{.3in}
      (\,\tilde{u}_2\,)_{\hat{\imath}}\,^j \,\,=\,\,
      \ba{cc}&-1\\0&\ea
      \nonumber\\[.1in]
      (\,\tilde{d}_1\,)_{\hat{\imath}}\,^j &=&
      \ba{cc}1&\\&0 \ea
      \hspace{.3in}
      (\,\tilde{d}_2\,)_{\hat{\imath}}\,^j \,\,=\,\,
      \ba{cc}&0\\1&\ea \,,
 \label{lmats1}\err
 where blank matrix entries represent zeros.
 This set of eight matrices is completely equivalent to the
 transformation rules (\ref{ex2}).  It is straightforward to
 verify, using (\ref{qqt}), that the algebra (\ref{q1alg}) is properly
 represented using these matrices.

 As an example, to illustrate what
 these matrices mean, consider the matrix $\tilde{u}_2$ defined in
 (\ref{lmats1}).  This is the ``second fermion up matrix", where the qualifier ``second"
 refers
 to the subscript on $\tilde{u}_2$ and indicates that this matrix encodes a mapping under the second supersymmetry,
 while the qualifier ``fermionic" refers to the tilde, and indicates that this matrix encodes transformations of the
 fermions.  The single non-vanishing term in this matrix is in
 the first row, second column, which indicates that, of the two
 fermions, only the first fermion $\psi_1$ transforms under the
 second supersymmetry, into the second boson
 $\phi_2$.  The fact that this matrix entry is $-1$ indicates a
 minus sign in the transformation rule $\delta_Q\,\psi_1$ on the
 term proportional to $\phi_2$, \ie, $\delta_Q\,\psi_1=\cdots-\phi_2\,\e^2$, as seen in (\ref{ex2}).
 The reason why this is called an ``up" matrix is that it encodes a
 mapping ``upward" from from a field with lower engineering
 dimension --- $\psi_1$ in this case--- to a field with higher
 engineering dimension --- $\phi_2$ in this case.

 The matrices in (\ref{lmats1}) exhibit the properties
 $\tilde{u}_A=d_A^T$ and $\tilde{d}_A=u_A^T$.  It is easy to see that this indicates a symmetric feature in the
 transformation rules (\ref{ex2}), whereby a fermion appearing in a
 boson transformation rule is correlated with that
 boson appearing in the transformation rule for that fermion.  In
 other words, terms in these transformation rules come paired.  This
 feature is satisfied by a wide and important class of supermultiplets, which
 we call ``standard".  (These are also called ``Adinkraic" in the
 literture.)

 There is a third equivalent way to represent the supersymmetry transformations
 given by (\ref{ex2}) and by (\ref{lmats1}).  This method uses the
 observation that the generic properties of linkage matrices
 facilitate a concise system under which the entire collection of
 linkage matrices for a given multiplet can be faithfully represented by a graph.
 Such a graph, called an Adinkra, consists of $d$ white
 vertices (one for each boson) and $d$ black vertices (one for each
 fermion).  Two vertices are connected by an $A$-th colored edge if
 the two fields corresponding to those vertices are inter-related by the $A$-th
 supersymmetry.  The edge is rendered solid if the corresponding
 $Q_A$ matrix entries are +1 and are rendered dashed if the
 corresponding $Q_A$ matrix entries are $-1$.  Finally, the vertices
 are arranged so that their heights on the graph correlate
 faithfully with the respective engineering dimension.

 Thus, if we designate $Q_1$ using {\purple purple} edges and $Q_2$
 using {\blue blue} edges, then the example multiplet described by
 (\ref{lmats1}), equivalent to (\ref{ex2}), would have the following
 Adinkra,
 \brr \includegraphics[width=1.6in]{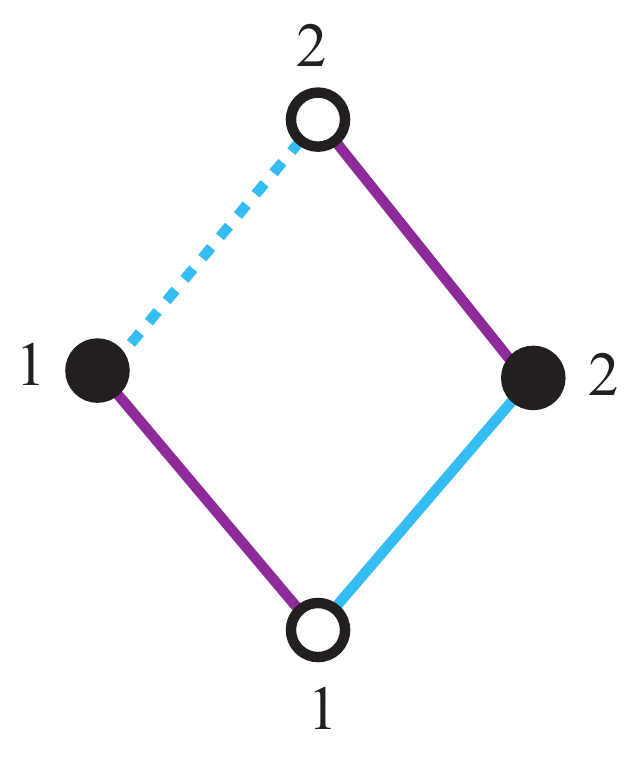}
 \label{exad}\err
 where the numerals on the vertices specify the fields, \eg, the
 black vertex labeled 2 represents the fermion field $\psi_2$.  As
 an easy exercise, the reader should confirm that (\ref{lmats1}) can
 be recovered from (\ref{exad}) using the rules described above.
 There is a striking economy exhibited by this graphical method,
 empowered by the fact that these graphs completely encode every aspect of the transformation
 rules, in a way which allows for ready translation from any Adinkra
 into linkage matrices or into parameter-dependent transformation
 rules.

 As another example, consider the following Adinkra,
 \brr \includegraphics[width=.8in]{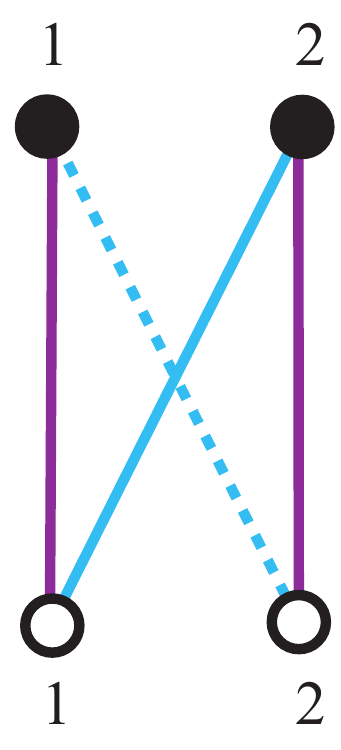}
 \label{adex2}\err
 This describes a supermultiplet distinct from the previous example,
 as evidenced by the fact that (\ref{adex2}) spans only two
 different engineering dimensions, whereas (\ref{exad}) spans three.

 We can readily extract the linkage matrices equivalent to (\ref{adex2}).
 For example, the boson down matrices $d_1$ and $d_2$ obviously
 vanish because the two bosons do not connect ``downward"
 to any lower fermion vertices.  Similarly, the two fermion up
 matrices $\tilde{u}_1$ and $\tilde{u}_2$ also obviously vanish, since there are no links ``upward" from the
 black vertices.  We can determine the non-vanishing linkage matrices by ``reading" the diagram.
 For example, the boson up matrix $u_2$ encodes blue edges connecting
 upward from boson vertices.  Thus, since the boson $\phi_1$
 links upward via blue edge only to the fermion $\psi_2$,
 and does so with a solid edge, this tells us that the matrix entry
 $(\,u_2\,)_1\,^2=+1$.  In this way, we can translate the Adinkra
 (\ref{adex2}) into the linkage matrices described by $d_A=0$,
 $\tilde{u}_A=0$,
  \brr (\,u_1\,)_i\,^{\hat{\jmath}} &=&
      \ba{cc} 1& \\ & 1 \ea
      \hspace{.4in}
      (\,u_2\,)_i\,^{\hat{\imath}} \,\,=\,\,
      \ba{cc}&1\\-1& \ea
       \,,
 \label{lmats2}\err
 and $\tilde{d}_A=u_A^T$.

 Note that the Adinkra (\ref{exad}) can be obtained from (\ref{adex2}) by
 an interesting operation: by moving the vertex $\phi_2$ to a new
 position located one level above the fermions, while continuously
 maintaining all inter-vertex edge connections, so that the edges
 swivel upward during this process.  This macrame-like move encodes a
 transformation which maps one supermultiplet into another, and is
 called a vertex raising operation.

 One of our results concerning Adinkras is a mathematical proof that
 any standard supermultiplet can be obtained by a sequence of vertex
 raising operations starting from an Adinkras with vertices which
 span only two different heights, \eg, (\ref{adex2}).  Accordingly,
 the representation theory of 1D standard multiplets breaks naturally
 into two parts; first to classify all of the possible two-height
 Adinkras for a given value of $N$, and then to systematize the
 possible sequences of vertex raises using each of
 these as a starting point.

 Owing to the special role played by the two-height Adinkras, we
 have given these a special name.  Standard Adinkras which span only
 two height assignments are called Valise Adinkras, or Valises for
 short.  The reason for this nomenclature is based on the observation that a
 large number of multiplets can be ``unpacked", as from a
 suitcase (or a valise), by judicious choices of vertex
 raises.\footnote{We credit Tristan H{\"u}bsch for inventing this
 catchy and useful term.}

 Using the information above, the reader ought to be able to verify
 the relationships between the Adinkras shown in Figures \ref{val4},
 \ref{a341a}, and \ref{a242a}, with the corresponding linkage
 matrices exhibited in the respective Tables \ref{links44}, \ref{links341}, and
 \ref{links242}, and should appreciate our use of the terms Adinkra,
 Valise, and the concept of vertex raising.

 \renewcommand{\theequation}{C.\arabic{equation}}
 \section{The shadow of the Chiral multiplet}
 \label{Chishadow}
 In this Appendix we explain how to dimensionally-reduce
 the 4D Chiral multiplet to extract its shadow.

 The 4D $N=1$ Chiral multiplet has the following transformation
 rules,
 \brr \delta_Q\,\phi &=& 2\,i\,\bar{\e}_L\,\chi_R
      \nonumber\\[.1in]
      \delta_Q\,\chi_R &=& \dslash\,\phi\,\e_L
      +F\,\e_R
      \nonumber\\[.1in]
      \delta_Q\,F &=&
      2\,i\,\bar{\e}_R\,\dslash\chi_R \,,
 \label{chiral}\err
 where $\phi$ is a complex scalar, $F$ is a complex auxiliary
 scalar, and $\chi_R$ is a right-chiral Weyl spinor field.
 The parameter $\e_R$ is also a right-chiral spinor, while
 $\e_L$ describes the corresponding Majorana conjugate, \ie,
 $\e_L=C^{-1}\,\bar{\e}_R^T$.
 The transformation rules (\ref{chiral}) satisfy $[\,\delta_Q(\e_1)\,,\,\delta_Q(\e_2)\,]
 =4\,i\,\bar{\e}_{[2\,L}\,\dslash\,\e_{1]\,L}$ on all component fields $\phi$, $F$,
 and $\chi_R$.  Note that we can define a Majorana spinor parameter
 via $\e=\e_R+\e_L$, so that $\e_{R,L}=\fr12\,(1\pm\Gamma_5)\,\e$ are the
 corresponding right- and left-chiral projections.  In terms of the
 Majorana spinor, the algebra is $[\,\delta_Q(\e_1)\,,\,\delta_Q(\e_2)\,]
 =2\,i\,\bar{\e}_{2}\,\dslash\,\e_{1}$.

 We express spinors in the Majorana basis described in Appendix
 \ref{spinbases}.\footnote{The choice of basis is immaterial for the computing the
 dimensional reduction; we obtain identical results using any other basis.
 We use the Majorana basis here in order to maintain consistency with
 other derivations in this paper.}
 Accordingly, we write the spinor field and the spinor supersymmetry parameter as
 \brr \chi_R &=&
      \fr12\,\ba{c}\chi_1+i\,\chi_2\\\chi_2-i\,\chi_1\\\hline
      \chi_3+i\,\chi_4\\\chi_4-i\,\chi_3 \ea
      \hspace{.3in}
      \e_R \,\,=\,\,
      \fr12\,\ba{c} \e_1+i\,\e_2 \\
      \e_2-i\,\e_1 \\\hline
      \e_3+i\,\e_4 \\
      \e_4-i\,\e_3 \ea \,,
 \label{spinchoice}\err
 where $\chi_{1,2,3,4}$ are each real anti-commuting fields
 and $\e_{1,2,3,4}$ are each real anti-commuting constant parameters.
 We also write the complex boson fields as $\phi=\phi_1+i\,\phi_2$, where
 $\phi_{1,2}$ are real bosons, and $F=F_1+i\,F_2$, where $F_{1,2}$ are
 real auxiliary bosons.

 \begin{figure}
 \begin{center}
 \includegraphics[width=2in]{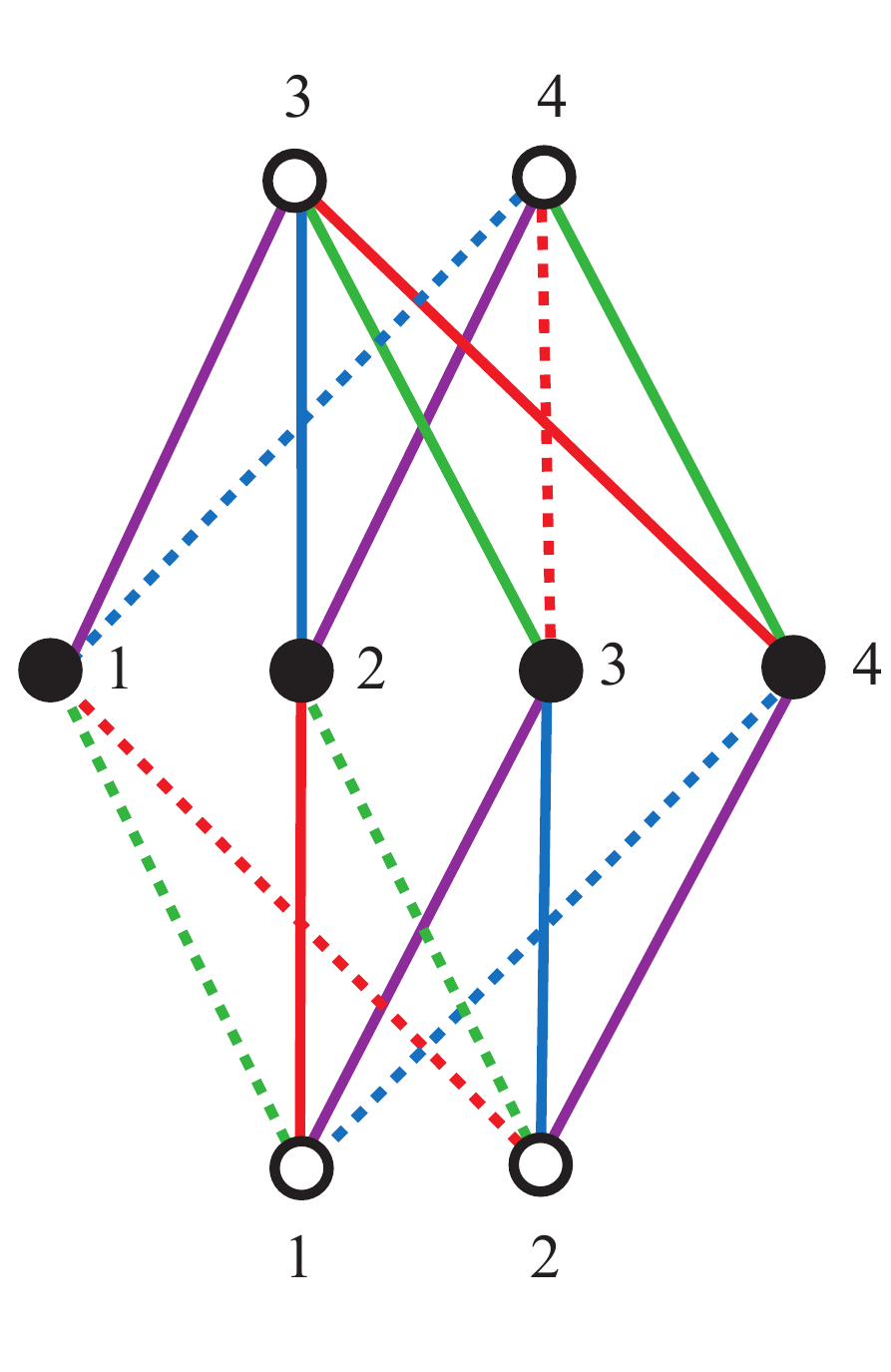}
 \caption{The Shadow of the Chiral multiplet, expressed as an Adinkra equivalent to the transformation rules
 (\ref{chshr}.) }
 \label{chshad}
 \end{center}
 \end{figure}

 Using these
 definitions, setting $\der_a=0$, and using the spinor identities in
 Appendix \ref{spinbases}, the
 transformation rules (\ref{chiral}) become
 \brr \delta_Q\,\phi_1 &=&
      -i\,\e_1\,\chi_3+i\,\e_2\,\chi_4+i\,\e_3\,\chi_1-i\,\e_4\,\chi_2
      \nonumber\\[.1in]
      \delta_Q\,\phi_2 &=&
      -i\,\e_1\,\chi_4-i\,\e_2\,\chi_3+i\,\e_3\,\chi_2+i\,\e_4\,\chi_1
      \nonumber\\[.1in]
      \delta_Q\,\chi_1 &=&
      F_1\,\e_1-F_2\,\e_2-\dot{\phi}_1\,\e_3-\dot{\phi}_2\,\e_4
      \nonumber\\[.1in]
      \delta_Q\,\chi_2 &=&
      F_2\,\e_1+F_1\,\e_2-\dot{\phi}_2\,\e_3+\dot{\phi}_1\,\e_4
      \nonumber\\[.1in]
      \delta_Q\,\chi_3 &=&
      \phi_1\,\e_1+\phi_2\,\e_2+F_1\,\e_3-F_2\,\e_4
      \nonumber\\[.1in]
      \delta_Q\,\chi_4 &=&
      \phi_2\,\e_1-\phi_1\,\e_2+F_2\,\e_3+F_1\,\e_4
      \nonumber\\[.1in]
      \delta_Q\,F_1 &=&
      -i\,\e_1\,\dot{\chi}_1-i\,\e_2\,\dot{\chi}_2-i\,\e_3\,\dot{\chi}_3-i\,\e_4\,\dot{\chi}_4
      \nonumber\\[.1in]
      \delta_Q\,F_2 &=&
      -i\,\e_1\,\dot{\chi}_2+i\,\e_2\,\dot{\chi}_1-i\,\e_3\,\dot{\chi}_4+i\,\e_4\,\dot{\chi}_3
      \,.
 \label{chshr}\err
 These rules describe the shadow of (\ref{chiral}).
 We organize the boson fields so that
 $\phi_i=(\,\phi_1\,,\,\phi_2\,,\,F_1\,,\,F_3\,)$ and the fermion
 fields so that
 $\psi_{\hat{\imath}}=(\,\chi_1\,,\,\chi_2\,,\,\chi_3\,,\,\chi_4\,)$.
 Using the Adinkra conventions described in Appendix \ref{adinkrastuff}, along with
 our edge coloration scheme whereby $Q_{1,2,3,4}$ are respectively described by {\purple
 purple}, {\blue blue}, {\green green}, and {\red red} colored
 edges, we can unambiguously represent (\ref{chshr}) as the Adinkra shown in Figure \ref{chshad}.

 It is easy to translate the Adinkra in Figure \ref{chshad} into equivalent up and down
 matrices.  For example, we determine the boson up matrix $u_1$ by
 looking at the purple colored edges extending upward from boson
 vertices.  There are two such edges: a solid edge connecting
 $\phi_1$ with $\psi_3$ and solid edge connecting $\phi_2$ with
 $\psi_4$.  Thus, there are two non-vanishing entries in $u_1$:
 one in the first row, third column, and the other in the second
 row, fourth column. These both take the value +1 because both edges are solid.

 \begin{table}
 \begin{center}
 \begin{tabular}{cc}
 $u_1 \,\,=\,\, \ba{cc|cc}&1&&\\-1&&&\\\hline &&&0\\&&0& \ea$ &
 $u_2 \,\,=\,\, \ba{cc|cc}&&&-1\\&&1& \\\hline &0&&\\0&&& \ea$
 \\[.5in]
 $u_3 \,\,=\,\, \ba{cc|cc}-1&&&\\&-1&& \\\hline &&0& \\&&&0 \ea$ &
 $u_4 \,\,=\,\, \ba{cc|cc}&1&&\\-1&&&\\\hline &&&0\\&&0& \ea$
 \end{tabular}
 \caption{The boson up linkage matrices for the Chiral multiplet shadow.  The fermion
 time-like down matrices are determined from these via $\tilde{d}_A=u_A^T$. }
 \label{chiralups}
 \vspace{1.1in}
 \begin{tabular}{cc}
 $\Delta^0_1 \,\,=\,\, \ba{cc|cc} &&0& \\&&&0 \\\hline 1 &&& \\&1&& \ea$ &
 $\Delta^0_2 \,\,=\,\, \ba{cc|cc} &&&0 \\&&0& \\\hline &1&&\\-1&&& \ea$
 \\[.5in]
 $\Delta^0_3 \,\,=\,\, \ba{cc|cc} 0&&& \\&0&& \\\hline &&1& \\&&&1 \ea$ &
 $\Delta^0_4 \,\,=\,\, \ba{cc|cc} &0&&\\0&&& \\\hline &&&1\\&&-1& \ea$
 \end{tabular}
 \caption{The time-like dowm linkage matrices for the Chiral
 multiplet. The fermion up matrices are determined from these via $\tilde{u}_A=d_A^T$. }
 \label{chiraldown0s}
 \end{center}
 \end{table}

 We can also determine the time-like down linkage matrices directly
 from (\ref{chiral}).  For example, to determine the femion 1-sector
 down matrices, $\tilde{\Delta}^1_A$ we isolate those terms in the
 fermion transformation rule $\delta_Q\,\psi_R$ involving the derivative
 $\der_1$.  These are given by
 $\delta_Q^{(1)}\,\chi_R=\Gamma^1\,\e_L\,\der_1\phi$.  We then use
 the explicit matrix $\Gamma^1$ specified in (\ref{gammaj}), the
 spinor components specified in (\ref{spinchoice}), and we write
 $\phi=\phi_1+i\,\phi_2$.  This allows us, after a small amount of algebra, to
 re-write the 1-sector
 fermion transformation rule as
 $\delta_Q\,\psi_{\hat{\imath}}=\e^A\,(\,\tilde{\Delta}^1_A\,)_{\hat{\imath}}\,^j\,\der_1\phi_j$,
 from which we can read off the four matrices
 $(\,\tilde{\Delta}^1_A\,)_{\hat{\imath}}\,^j$.  It is
 straightforward to perform this calculation, and then to verify
 that the matrices thereby obtained satisfy
 $\tilde{\Delta}^1_A=-(\,\Gamma^0\Gamma^1\,)_A\,^B\,\tilde{\Delta}^0_B$,
 where $\tilde{\Delta}^0_A=u_A^T$.  Similar calculations can be done
 for all of the time-like down matrices, providing a nice
 consistency check on our powerful assertion (\ref{ga}).

 \renewcommand{\theequation}{D.\arabic{equation}}
 \section{The shadow of the Maxwell field strength multiplet}
 \label{maxlinks}
 In this Appendix we determine the linkage matrices
 for the 4D $N=4$ Maxwell field strength multiplet.  This provides a
 means to exhibit precisely how 1D phantom sectors arise upon
 restriction of a $p=1$ gauge multiplet to a zero-brane.  This
 Appendix is complementary to section \ref{phan} in the main text,
 above.

 The 4D $N=1$ super-Maxwell field-strength multiplet has the
 following transformation rules,
 \brr \delta_Q\,\lambda &=&
      \fr12\,F_{\mu\nu}\,\Gamma^{\mu\nu}\,\e
      -i\,D\,\Gamma_5\,\e
      \nonumber\\[.1in]
      \delta_Q\,F_{\mu\nu} &=&
      -2\,i\,\bar{\e}\,\Gamma_{[\mu}\,\der_{\nu]}\,\lambda
      \nonumber\\[.1in]
      \delta_Q\,D &=&
      \bar{\e}\,\Gamma_5\,\dslash\,\lambda \,,
 \label{maxrules}\err
 where $\lambda$ is a Majorana spinor gaugino field, $D$ is a real auxiliary (pseudo)scalar, and
 the field strength tensor $F_{\mu\nu}$ is subject to the Bianchi identity
 $\der_{[\lambda}F_{\mu\nu]}=0$.
 We obtain the linkage matrices equivalent to (\ref{maxrules})
 by de-constructing these rules using a specific spinor basis, and re-writing
 them in terms of individual degrees of freedom as specified in
(\ref{ssm}).
 It follows simply that that $\tilde{\Delta}^a_A=0$ and
 $u_A=0$, since the fermions $\lambda_A$ share a common engineering dimension of 3/2 while the bosons
 $F_{\mu\nu}$ and $D$ share a common engineering dimension of 2.

 We use the specific Majorana basis defined in Appendix \ref{spinbases} by the Gamma
 matrices given in (\ref{gammaj}).  To determine the ``up" linkage
 matrices, it is helpful to re-write the fermion transformation
 rule in (\ref{maxrules}) as
 \brr \delta_Q\,\lambda &=&
      2\,E_a\,{\cal B}^a\,\e
      +2\,B^a\,{\cal R}_a\,\e
      -i\,D\,\Gamma_5\,\e \,,
 \label{lamon}\err
 where we have used the definitions $E_a=F_{0a}$ and
 $B^a=\fr12\,\ve^{abc}\,F_{bc}$, for the electric and magnetic fields,
 respectively.  We have also used the definitions ${\cal
 B}_a=\fr12\,\Gamma^0\Gamma^a$, and ${\cal
 R}_a=\fr14\,\ve_{abc}\,\Gamma^{bc}$ for the boost and rotation
 generators also given in Appendix \ref{spinbases}.
 We now use the explicit matrices ${\cal B}_a$, ${\cal R}^a$, and $\Gamma_5$
 specified in (\ref{boorots}) and
 (\ref{gammaj}), to re-cast (\ref{lamon}) in matrix form:
 the left side as a four-component column matrix
 $\lambda_A=(\,\lambda_1\,,\,\lambda_2\,|\,,\,\lambda_3\,,\,\lambda_4\,)^T$
 and the right-hand side as a $4\times 7$ matrix multiplying another
 four-component column matrix given by
 $\e_A=(\,\e_1\,,\,\e_2\,|\,\e_3\,,\,\e_4\,)^T$.  A small amount of
 algebra then allows us to re-write the result in the form
 $\delta_Q\,\lambda_{\hat{\imath}}=\e^A\,(\,\tilde{u}_A\,)_{\hat{\imath}}\,^j\,\phi_j$,
 where $\phi_i:=(\,E_1\,,\,E_2\,|\,E_3\,,\,D\,||\,B_1\,,\,B_2\,,\,B_3\,)^T$,
  whereupon we can read off each of the four matrices
 $(\,\tilde{u}_A\,)_{\hat{\imath}}\,^j$.  The result is shown in
 Table \ref{maxups}.  The fact that $\lambda$ transforms
 non-trivially into $B^a$ manifests in the non-triviality of the
 rightmost three columns in these results.

 We then do a similar thing to the boson fields to determine the
 down matrices $\Delta^\mu_A$.  We do this separately for each
 of the four choices for $\mu$, referring to these as the
 $\mu$-sector down matrices.  For example, to extract the 0-sector
 down matrices, we isolate those terms in the boson transformation
 rules in (\ref{maxrules}) proportional to the derivative
 $\der_0\lambda$.  These are given by
 \brr \delta_Q^{(0)}\,E_a &=& i\,\bar{\e}\,\Gamma_a\,\der_0\lambda
      \nonumber\\[.1in]
      \delta_Q^{(0)}\,D &=& \bar{\e}\,\Gamma_5\Gamma^0\,\der_0\lambda
      \nonumber\\[.1in]
      \delta_Q^{(0)}\,B^a &=& 0 \,.
 \label{d0b}\err
 Note that the magnetic fields
 $B^a=\fr12\,\ve^{abc}\,F_{bc}$ do not transform into time
 derivatives of the gaugino field.  This is not surprising since the
 magnetic field is expressible locally as
 $B^a=\ve^{abc}\,\der_bA_c$.  But it is worth noting that
 (\ref{d0b}) follows simply from (\ref{maxrules}).  This tells us
 that upon restriction to a zero-brane, there are no downward
 Adinkra links connecting the three magnetic field components to any
 other fields; in the shadow these degrees of freedom sit at the top
 of one-way upward edges.  In this way the magnetic fields de-couple
 from the multiplet upon reduction to one-dimension.  By utilizing
 the specific Gamma matrices given in Appendix \ref{spinbases} we
 can use the same techniques described above to re-write the
 0-sector transformation rules (\ref{d0b}) as
 $\delta_Q\,\phi_i=-i\,\e^A\,(\,\Delta^0_A\,)_i\,^{\hat{\imath}}\,\der_0\lambda_{\hat{\imath}}$,
 and then read-off the the matrices
 $(\,\Delta^0_A\,)_i\,^{\hat{\imath}}$.  The result of this
 straightforward process is exhibited in Table \ref{maxdown0}.\footnote{It is easy
 to see that $\tilde{u}_A^T\ne\Delta^0_A$, so that in this case the Phantom
 matrix defined in (\ref{phantasm}) is non-vanishing.  Although the
 phantom sector is irrelevant to any one-dimensional
 physics, it is necessary to resurrect this sector should we wish
 to enhance the shadow theory to its full ambient analog.}

 By isolating the terms in (\ref{maxrules})
 respectively proportional to $\der_{1,2,3}\lambda$, writing these explicitly using the
 Majorana basis Gamma matrices shown in (\ref{gammaj}), and then
 re-configuring the rules as
 $\delta_Q\,\phi_i=-i\,\e^A\,(\,\Delta^a_A\,)_i\,^{\hat{\imath}}\,\der_a\lambda_{\hat{\imath}}$,
 allows us to read off the remaining space-like linkage matrices.
 The results of this straightforward process are exhibited in Tables
 \ref{maxdown1}, \ref{maxdown2}, and \ref{maxdown3}.

 \begin{table}
 \begin{center}
 \begin{tabular}{cc}
 $\tilde{u}_1=\ba{cc|cc||ccc}
      &&1&&0&0&0 \\&&&-1&0&0&1\\\hline
      1&&&&0&-1&0 \\&1&&&1&0&0 \ea$ &
 $\tilde{u}_2=\ba{cc|cc||ccc}
      &&&1&0&0&-1 \\&&1&&0&0&0 \\\hline
      &-1&&&-1&0&0 \\1&&&&0&-1&0 \ea$ \\[.6in]
 $\tilde{u}_3=\ba{cc|cc||ccc}
      1&&&&0&1&0 \\&-1&&&1&0&0 \\\hline
      &&-1&&0&0&0 \\ &&&-1&0&0&-1 \ea$ &
 $\tilde{u}_4=\ba{cc|cc||ccc}
      &1&&&-1&0&0 \\ 1&&&&0&1&0 \\\hline
      &&&1&0&0&1 \\ &&-1&&0&0&0 \ea$
 \end{tabular}
 \caption{The four ``up" linkage matrices associated with the Maxwell field strength
 multiplet.}
 \vspace{.7in}
 \label{maxups}
  \begin{tabular}{cc}
 $\Delta^0_1 \,\,=\,\, \ba{cc|cc}&&1&\\&&&1\\\hline
      1&&&\\&-1&&\\\hline\hline
      0&0&0&0\\0&0&0&0\\0&0&0&0\ea$ &
 $\Delta^0_2 \,\,=\,\, \ba{cc|cc}&&&1\\&&-1&\\\hline
      &1&&\\1&&&\\\hline\hline
      0&0&0&0\\0&0&0&0\\0&0&0&0\ea$ \\[1in]
 $\Delta^0_3 \,\,=\,\, \ba{cc|cc}1&&&\\&-1&&\\\hline
      &&-1& \\&&&-1 \\\hline\hline
      0&0&0&0\\0&0&0&0\\0&0&0&0\ea$ &
 $\Delta^0_4 \,\,=\,\, \ba{cc|cc}&1&&\\1&&&\\\hline
      &&&-1\\&&1& \\\hline\hline
      0&0&0&0\\0&0&0&0\\0&0&0&0\ea$
 \end{tabular}
 \caption{The time-like down matrices associated with the Maxwell
 field strength multiplet.}
 \label{maxdown0}
 \end{center}
 \end{table}

 \begin{table}
 \begin{center}
 \begin{tabular}{cc}
 $\Delta^1_1 \,\,=\,\, \ba{cc|cc}-1&&&\\&0&&\\\hline
      &&0&\\&&&1\\\hline\hline
      0&0&0&0\\-1&0&0&0\\0&0&0&1\ea$ &
 $\Delta^1_2 \,\,=\,\, \ba{cc|cc}&-1&&\\0&&&\\\hline
      &&&0\\&&-1& \\\hline\hline
      0&0&0&0\\0&-1&0&0\\0&0&-1&0\ea$ \\[1in]
 $\Delta^1_3 \,\,=\,\, \ba{cc|cc} &&-1&\\&&&0\\\hline
      0&&&\\&1&& \\\hline\hline
      0&0&0&0\\0&0&1&0\\0&-1&0&0\ea$ &
 $\Delta^1_4 \,\,=\,\, \ba{cc|cc} &&&-1\\&&0& \\\hline
      &0&&\\-1&&& \\\hline\hline
      0&0&0&0\\0&0&0&1\\1&0&0&0\ea$
 \end{tabular}
 \caption{The 1-sector space-like down matrices for the Maxwell field strength
 multiplet.}
 \label{maxdown1}
 \vspace{.4in}
 \begin{tabular}{cc}
 $\Delta^2_1 \,\,=\,\, \ba{cc|cc} &0&&\\-1&&&\\\hline
      &&&0\\&&-1& \\\hline\hline
      1&0&0&0\\0&0&0&0\\0&0&-1&0\ea$ &
 $\Delta^2_2 \,\,=\,\, \ba{cc|cc} 0&&&\\&-1&&\\\hline
      &&0& \\&&&-1\\\hline\hline
      0&1&0&0\\0&0&0&0\\0&0&0&-1\ea$ \\[1in]
 $\Delta^2_3 \,\,=\,\, \ba{cc|cc} &&&0\\&&-1& \\\hline
      &0&&\\1&&& \\\hline\hline
      0&0&-1&0\\0&0&0&0\\-1&0&0&0\ea$ &
 $\Delta^2_4 \,\,=\,\, \ba{cc|cc} &&0& \\&&&-1 \\\hline
      0&&&\\&1&& \\\hline\hline
      0&0&0&-1\\0&0&0&0\\0&-1&0&0\ea$
 \end{tabular}
 \caption{The 2-sector space-like down matrices for the Maxwell
 field strength multiplet.}
 \label{maxdown2}
 \end{center}
 \end{table}

 \begin{table}
 \begin{center}
 \begin{tabular}{cc}
 $\Delta^3_1 \,\,=\,\, \ba{cc|cc} &&0& \\&&&0 \\\hline
      -1&&&\\&1&& \\\hline\hline
      0&0&0&-1\\0&0&1&0\\0&0&0&0\ea$ &
 $\Delta^3_2 \,\,=\,\, \ba{cc|cc} &&&0 \\ &&0& \\\hline
      &-1&&\\-1&&& \\\hline\hline
      0&0&1&0\\0&0&0&1\\0&0&0&0\ea$ \\[1in]
 $\Delta^3_3 \,\,=\,\, \ba{cc|cc} 0&&&\\&0&& \\\hline
      &&-1& \\&&&-1 \\\hline\hline
      0&1&0&0\\1&0&0&0\\0&0&0&0\ea$ &
 $\Delta^3_4 \,\,=\,\, \ba{cc|cc} &0&& \\0&&& \\\hline
      &&&-1 \\&&1& \\\hline\hline
      -1&0&0&0\\0&1&0&0\\0&0&0&0\ea$
 \end{tabular}
 \caption{The 3-sector space-like down matrices for the Maxwell
 field strength multiplet.}
 \label{maxdown3}
 \end{center}
 \end{table}

 \renewcommand{\theequation}{E.\arabic{equation}}
 \section{4D Spinor bases}
 \label{spinbases}
 Gamma matrices satisfy the Clifford relationship
 $\{\,\Gamma_\mu\,,\,\Gamma_\nu\,\}_A\,^B=-2\,\eta_{\mu\nu}\,\delta_A\,^B$, where
 $\eta_{\mu\nu}={\rm diag}(+---)$.  These act from the left on spinors
 $\psi_A$ and from the right on barred spinors
 $\bar{\psi}^A:=(\,\psi^\dagger\,\Gamma_0\,)^A$.
 In four-dimensions, the minimal solution involves $4\times 4$
 matrices, so the spinor index $A$ takes on four values.
 The 4D charge conjugation matrix $C$ is
 defined by $C\,\Gamma^a\,C^{-1}=-(\,\Gamma^a\,)^T$.  In addition,
 the matrix $C$ is real, antisymmetric, and has unit
 determinant.  A chirality operator
 is defined by $\Gamma_5:=i\,\Gamma_0\Gamma_1\Gamma_2\Gamma_3$.

 We can change bases by replacing
 \brr \Psi_A &\to& M_A\,^B\,\psi_B
      \nonumber\\[.1in]
      (\,\Gamma^\mu\,)_A\,^B &\to&
      (\,M\,\Gamma^\mu\,M^{-1}\,)_A\,^B
      \nonumber\\[.1in]
      C^{-1}_{AB} &\to& \frac{1}{\sqrt{\det(M)}}\,(\,M\,C^{-1}\,M^T\,)_{AB} \,.
 \label{bchange}\err
 where $M$ is any nonsingular $4\times 4$ matrix.   Two especially useful bases are the
 Weyl basis and the Majorana basis.  These are explained below.
 Note that $G^\mu=-\Gamma^\mu\,C^{-1}$ transforms as
 \brr G^\mu &\to&
      \frac{1}{\sqrt{\det(M)}}\,M\,G^\mu\,M^T
 \err
 Given any basis, this allows us to find a similarity transformation
 to render all spinor components real (a Majorana basis), and
 moreover one for which $G^0_{AB}=\delta_{AB}$.
 The resultant basis is then specially tailored
 for dimensional reduction to 1D, for the simple reason that
 the four real components of the Majorana supercharge operator $Q_A$
 supply natural real 1D shadow supercharges, which satisfy the
 algebra $\{\,Q_A\,,\,Q_B\,\}=i\,\delta_{AB}\,\der_\tau$.

 \subsection{The Weyl basis}
 In the Weyl basis we choose
 $4\times 4$ matrices using the following convention,
 \brr \Gamma_0 &=& \ba{c|c}&-\Ione\\\hline \Ione & \ea
      \hspace{.4in}
      \Gamma_a \,\,=\,\,
      \ba{c|c}&\sigma_a\\\hline \sigma_a&\ea
      \nonumber\\[.1in]
      C &=& \ba{c|c}\ve&\\\hline&\ve\ea
      \hspace{.5in}
      \Gamma_5 \,\,=\,\,
      \ba{c|c}\Ione&\\\hline&-\Ione\ea \,,
 \label{gammats}\err
 where $\Ione$ is the $2\times 2$ unit matrix, $a=1,2,3$, and $\sigma_a$ are the Pauli
 matrices and $\ve=i\,\sigma_2$.

 Right- and left-handed Weyl spinors satisfy the respective constraints
 $\Gamma_5\,\psi_{R,L}=\pm \psi_{R,L}$.  In terms of the Weyl basis
 (\ref{gammats}), this means that right- and left-handed spinors are respectively configured as
 \brr \chi_R &=& \ba{c} \chi_1 \\\chi_2 \\\hline 0\\0 \ea
      \hspace{.3in}
      \varphi_L \,\,=\,\, \ba{c} 0 \\0\\\hline \varphi_1 \\\varphi_2
      \ea \,,
 \err
 where $\chi_1$, $\chi_2$, $\varphi_1$ and $\varphi_2$ are complex
 anticommuting fields.
 Note that Weyl spinors take an especially tidy form in the Weyl
 basis, since half of the four complex spinor components vanish.

 A Majorana spinor satisfies $\psi=C^{-1}\,\bar{\psi}^T$.  In terms
 of the Weyl basis (\ref{gammats}), this means
 \brr \psi &=&
      \ba{c} \psi_1 \\\psi_2\\\hline \psi_2^*\\-\psi_1^* \ea \,,
 \label{maj}\err
 where $\psi_1$ and $\psi_2$ are complex anticommuting fields.
 Note that Majorana spinors are relatively awkward in the Weyl basis.

 \subsection{The Majorana basis}
 \label{mbasis}
 Change bases from the Weyl basis to the Majorana basis,
 using (\ref{bchange}), by choosing
 \brr M \,\,=\,\,
      \fr12\,\ba{cc|cc} 1 & & & -1 \\ -i & & & -i \\\hline
      & 1 & 1 & \\
      & -i & i & \ea \,.
 \err
 Using the transformation (\ref{bchange}), right- and left-handed Weyl spinors in the Weyl
 basis transform into right- and left-handed Weyl spinors in the Majorana basis, as
 specified respectively by
 \brr \chi_{R,L} &=&
      \ba{c}\chi_1\\\mp i\,\chi_1 \\\hline
      \chi_2\\ \mp i \chi_2 \ea\,,
 \label{msw}\err
 where $\chi_1$ and $\chi_2$ are complex fields.
 Note that the difference between left- and right-handedness in this
 basis manifests in the relative phases appearing in (\ref{msw}).
 Note that Weyl spinors are relatively awkward in the Majorana
 basis.

 Using the transformation (\ref{bchange}), a Majorana
 spinor in the Weyl basis, (\ref{maj}), transforms into a
 Majorana spinor in the Majorana basis, as given by
 \brr \psi_A &=&
      \ba{c} {\rm Re}\,\psi_1 \\
      {\rm Im}\,\psi_1 \\\hline
      {\rm Re}\,\psi_2 \\
      {\rm Im}\,\psi_2 \ea \,.
 \err
 Note that Majorana spinors take an especially tidy form in the
 Majorana basis, since all four components are independent and real.
 In this basis, the Gamma matrices and the
 charge conjugation matrices are
 \brr \Gamma_0 &=&
      \ba{cc|cc}&&-1&\\&&&1\\\hline 1&&&\\&-1&& \ea
      \hspace{.3in}
      \Gamma_1 \,\,=\,\,
      \ba{cc|cc}-1&&&\\&1\\\hline &&1&\\&&&-1 \ea
      \nonumber\\[.1in]
      \Gamma_2 &=&
      \ba{cc|cc} &1&&\\1&&&\\\hline &&&1\\&&1&\ea
      \hspace{.5in}
      \Gamma_3 \,\,=\,\,
      \ba{cc|cc} &&1&\\&&&-1 \\\hline 1&&&\\&-1&& \ea
      \nonumber\\[.1in]
      C &=& \ba{cc|cc} &&1& \\&&&-1\\\hline -1&&&\\&1&& \ea
      \hspace{.3in}
      \Gamma_5 \,\,=\,\, \ba{cc|cc} &i&& \\-i&&&\\\hline &&&i\\&&-i&
      \ea \,,
 \label{gammaj}\err
 as obtained by transforming (\ref{gammats}) using (\ref{bchange}).
 The corresponding $G$-Matrices $G^a_{AB}=-\eta^{ab}\,(\,\Gamma_b\,C^{-1}\,)_{AB}$ are
 \brr G^0 &=& \ba{cc|cc}1&&&\\&1&&\\\hline &&1&\\&&&1\ea
      \hspace{.3in}
      G^1 \,\,=\,\, \ba{cc|cc} &&1&\\&&&1\\\hline 1&&&\\&1&& \ea
      \nonumber\\[.1in]
      G^2 &=& \ba{cc|cc} &&&1\\&&-1&\\\hline &-1&&\\1&&&\ea
      \hspace{.3in}
      G^3 \,\,=\,\, \ba{cc|cc}1&&&\\&1&&\\\hline &&-1& \\&&&-1 \ea
      \,.
 \label{Gmats}\err
 Note that $G$-matrices are symmetric, real, and traceless.

 Also useful are the ``boost" operators ${\cal B}^a=\fr12\,\Gamma^0\Gamma^a$ and the ``rotation" operators
 ${\cal R}_a=\fr14\,\ve_{abc}\,\Gamma^{bc}$. In the
 Majorana basis (\ref{gammaj}) these are
 \brr {\cal B}^1 &=& \fr12\,\ba{cc|cc}&&1&\\&&&1\\\hline 1&&&\\&1&& \ea
      \hspace{.6in}
      {\cal R}_1 \,\,=\,\, \fr12\,\ba{cc|cc}&&&-1\\&&1&\\\hline &-1&&\\1&&&\ea
      \nonumber\\[.1in]
      {\cal B}^2 &=& \fr12\,\ba{cc|cc}&&&1\\&&-1& \\\hline &-1&& \\1&&& \ea
      \hspace{.3in}
      {\cal R}_2 \,\,=\,\, \fr12\,\ba{cc|cc} &&1& \\&&&1\\\hline -1&&& \\&-1&& \ea
      \nonumber\\[.1in]
      {\cal B}^3 &=& \fr12\,\ba{cc|cc}1&&&\\&1&&\\\hline &&-1& \\&&&-1 \ea
      \hspace{.3in}
      {\cal R}_3 \,\,=\,\, \fr12\,\ba{cc|cc}&-1&&\\1&&&\\\hline &&&1\\&&-1&
      \ea \,.
 \label{boorots}\err
 Note that the boost operators are symmetric while the rotation
 operators are antisymmetric.  Note too that $G^a=2\,{\cal B}^a$.
 The operators in (\ref{boorots}) satisfy the Lorentz algebra
 \brr [\,{\cal R}_a\,,\,{\cal R}_b\,] &=& -\ve_{ab}\,^c\,{\cal R}_c
      \nonumber\\[.1in]
      [\,{\cal B}^a\,,\,{\cal B}^b\,] &=& \ve^{abc}\,{\cal R}_c
      \nonumber\\[.1in]
      [\,{\cal B}^a\,,\,{\cal R}_b\,] &=& -\ve^a\,_{bc}\,{\cal B}^c \,.
 \label{loren}\err
 The Lorentz algebra (\ref{loren}) can be written concisely, and in a manner which is
 manifestly covariant, as
 \brr [\,M_{\mu\nu}\,,\,M^{\lambda\sigma}\,] &=&
      \delta_\mu\,^\lambda\,M_\nu\,^\sigma
      -\delta_\mu\,^\sigma\,M_\nu\,^\lambda
      +\delta_\nu\,^\sigma\,M_\mu\,^\lambda
      -\delta_\nu\,^\lambda\,M_\mu\,^\sigma \,,
 \err
 where $M^{0a}={\cal B}^a$ and $M_{ab}=\ve_{abc}\,{\cal R}^c$.\footnote{Note that using our conventions,
 $M_{0a}=\eta_{00}\,\eta_{ab}\,M^{0b}=-\delta_{ab}\,M^{0b}$, whereas
 ${\cal B}_a=\delta_{ab}\,{\cal B}^b$.}

 \vspace{.1in}

 \noindent
 {\bf Acknowledgements}\\[.1in]
 The authors are thankful to their
 long-time collaborators Tristan H{\"u}bsch,
 Charles Doran, and S. J. Gates, Jr., for many relevant discussions which spawned this
 analysis.  M.F. is thankful also to his muse, Adriana, and to the the Slovak Institute for Basic
 Research (SIBR), in Podva${\check{\rm z}}$ie Slovakia, where much of this work was completed,
 for gracious hospitality.  The authors also thank the President's office at the
 SUNY College at Oneonta, for providing hospitable use of a campus
 guest house, where the ideas for this paper congealed.

 \Refs{References}{[00]}
 \bibitem{SS}
 J.~Wess and J.~Bagger,
 {\em Supersymmetry and supergravity},
 Princeton, USA: Univ. Pr. (1992);
  P.~Van Nieuwenhuizen,
  {\em Supergravity},
  Phys.\ Rept.\  {\bf 68}, 189 (1981);
 J.~Polchinski,
 {\em Supersymmetry And Supergravity};
 S.~Weinberg,
 {\em The quantum theory of fields.  Vol. 3: Supersymmetry},
 Cambridge, UK: Univ. Pr. (2000);
  M.~F.~Sohnius,
  {\it Introducing Supersymmetry},
  Phys.\ Rept.\  {\bf 128}, 39 (1985);
  S.~J.~Gates, M.~T.~Grisaru, M.~Rocek and W.~Siegel,
  {\em Superspace, or one thousand and one lessons in
  supersymmetry},
  Front.\ Phys.\  {\bf 58}, 1 (1983);
 \bibitem{QFT}
  S.~Weinberg:
  {\em The Quantum theory of fields. Vol. 1: Foundations}
  Cambridge, UK: Univ. Pr. (1995);
  S.~Weinberg:
  {\em The quantum theory of fields. Vol. 2: Modern applications},
  Cambridge, UK: Univ. Pr. (1996);
  M.~E.~Peskin and D.~V.~Schroeder:
  {\em An Introduction To Quantum Field Theory},
  Reading, USA: Addison-Wesley (1995);
   L.~H.~Ryder:
  {\em Quantum Field Theory},
  Cambridge, Uk: Univ. Pr. (1985);
  A.~Zee:
  {\em Quantum field theory in a nutshell},
  Princeton, UK: Princeton Univ. Pr. (2003);
  B.~De Wit and J.~Smith:
  {\em Field Theory In Particle Physics. Vol. 1},
  Amsterdam, Netherlands: North-holland (1986);
  \bibitem{Diverse}
  A.~Salam and E.~Sezgin:
  {\em Supergravities in diverse dimensions, vol 1,2},
 Singapore: World Scientific (1989);
 \bibitem{CJS}
  E.~Cremmer, B.~Julia and J.~Scherk:
  {\em Supergravity theory in 11 dimensions},
  Phys.\ Lett.\  B {\bf 76}, 409 (1978);
 \bibitem{GSW}
  M.~B.~Green, J.~H.~Schwarz and E.~Witten:
  {\em Superstring theory, vol. 1, 2},
  Cambridge, Uk: Univ. Pr. (1987);
  \bibitem{KK}
  T.~Kaluza:
 {\em Zum Unitätsproblem in der Physik},
 Sitzungsber. Preuss. Akad. Wiss. Berlin. (Math. Phys.) 1921;
 O.~Klein:
 {\em Quantentheorie und fünfdimensionale Relativitätstheorie},
 Zeitschrift für Physik a Hadrons and Nuclei 37 (12) 895;
 innumerable recent references;
 \Bib{FG1} M.~Faux and S.~J.~Gates, Jr.:
 {\em Adinkras: A Graphical Technology for Supersymmetric Representation Theory},
  Phys.~Rev.~{\bf D71}~(2005),~065002;
  \Bib{DFGHIL01} C.~Doran, M.~Faux, S.~J.~Gates, Jr., T.~H{\"u}bsch, K.~Iga, G.~Landweber:
 {\em On Graph Theoretic Identifications of
 Adinkras, Supersymmetry Representations and Superfields},
  Int. J. Mod. Phys. {\bf A}22 (2007) 869-930;
  \Bib{Prepotentials}
  C.~Doran, M.~Faux, S.~J.~Gates, Jr., T.~H{\"u}bsch, K.~Iga, G.~Landweber:
  {\em Adinkras and the Dynamics of Superspace Prepotentials},
  Adv. S. Th. Phys., Vol. 2, no. 3 (2008) 113-164;
  \Bib{Counter}
  C.~Doran, M.~Faux, S.~J.~Gates, Jr., T.~H{\"u}bsch, K.~Iga,
  G.~Landweber:
  {\em A counter example to a putative classification of
  one-dimensional $N$-extended supermultiplets},
  Advanced Studies in Theoretical Physics, Vol. 2, no. 3 (2008)
  99-111;
 \Bib{RETM}
 C.~Doran, M.~Faux, S.~J.~Gates, Jr., T.~H{\"u}bsch, K.~Iga, G.~Landweber:,
 {\it On the matter of $N=2$ matter},
 Phys. Lett. {\bf B}659 (2008) 441-446 \,;
  \Bib{Zeeman}
  C.~Doran, M.~Faux, S.~J.~Gates, T.~Hubsch, K.~Iga and
  G.~Landweber:
  {\em Super-Zeeman Embedding Models on N-Supersymmetric
  World-Lines},
   J.~Phys. {\bf A}42 (2009) 065402;
 \Bib{Frames}
 C.~Doran, M.~Faux, S.~J.~Gates, Jr., T.~H{\"u}bsch, K.~Iga, G.~Landweber:
 {\em Frames for supersymmetry},
  Int. J. Mod. Phys. {\bf A}24 (2009) 2665-2676;
  \Bib{AT1}
  C.~Doran, M.~Faux, S.~J.~Gates, Jr., T.~H{\"u}bsch, K.~Iga,
  G.~Landweber, and R.~L.~Miller:
  {\em Topology types of Adinkras and the corresponding representations of $N$-extended
  supersymmetry},
  arXiv:0806.0050;
  \Bib{HDS}
   C.~Doran, M.~Faux, S.~J.~Gates, Jr., T.~H{\"u}bsch, K.~Iga,
  G.~Landweber, and R.~L.~Miller:
  {\em Relating doubly-even error-correcting codes, graphs, and
  irreducible representations of $N$-extended supersymmetry},
  arXiv:0806.0051;
 \Bib{Codes}
 F.~J.~MacWilliams and N.~J.~A.~Sloane:
 {\em The theory of error-correcting codes},
 North Holland, Amsterdam, 1977 \,;
 J.~H.~Conway, V.~Pless and N.~J.~A.~Sloane:
 {\em The binary self-dual codes of length up to 32: a revised enumeration},
 J.~Combinatorial Theory A60 (1992) 183-195 \,;
 W.~C.~Huffman and V.~Pless:
 {\em Fundamentals of error-correcting codes},
 Cambridge University Press, 2003;
 \bibitem{enuf}
 S. J. Gates, Jr., W.D. Linch, III and
 J. Phillips:
 {\em When Superspace is Not Enough},
 hep-th/0211034;
   \Bib{T1}
   A.~Pashnev and F.~Toppan:
   {\em On the classification of $N$-extended supersymmetric quantum
   mechanical systems},
   J.~Math Phys. 42 (2001) 5257-5271;
   \Bib{T2}
   Z.~Kuznetsova, M.~Rojas, F.~Toppan,
   {\em Classification of irreps and invariants of the $N$-extended
   supersymmetric quantum mechanics},
   JHEP 03 (2006) 098;
   \Bib{T3}
   S.~Bellucci, S.~Krivonos, A.~Marrani, E.~Orazi:
   {\rm `Root' action for $N=4$ supersymmetric mechanics theories},
   Phys.~Rev.~D 73 (2006) 025011;
  \bibitem{T4}
  M.~Gonzales, Z.~Kuznetsova, A.~Nersessian, F.~Toppan and V.~Yeghikyan,
  {\em Second Hopf map and supersymmetric mechanics with Yang
  monopole},
  arXiv:0902.2682 [hep-th].
 \bibitem{T5}
  F.~Toppan:
  {\em Cliffordized NAC supersymmetry and PT-symmetric
  Hamiltonians},
  Fortsch.\ Phys.\  {\bf 56}, 516 (2008).
 \bibitem{T6}
  M.~Gonzales, M.~Rojas and F.~Toppan:
  {\em One-dimensional sigma-models with N=5,6,7,8 off-shell
  supersymmetries},
  arXiv:0812.3042 [hep-th].
 \bibitem{T7}
  Z.~Kuznetsova and F.~Toppan,
  Mod.\ Phys.\ Lett.\  A {\bf 23}, 37 (2008);
 \bibitem{FS1}
 M. Faux and D. Spector,
 {\em Duality and central charges in supersymmetric quantum
 mechanics}, Phys. Rev. {\bf D}70 (2004) 085014;
 \bibitem{FKS}
 M. Faux, D. Kagan, and D. Spector,
 {\em Central charges and extra dimensions in supersymmetric quantum
  mechanics},
  arXiv:hep-th/0406152;
 \bibitem{GatesRana1}
 S.J. Gates, Jr. and L. Rana,
 {\em A Theory of Spinning Particles for Large N Extended
 Supersymmetry},
 Phys.~Lett.{\bf B352} (1995) 50-58,
 hep-th/9504025\,.
 \bibitem{GatesRana2}
 S.J. Gates, Jr. and L. Rana,
 ``A Theory of Spinning Particles for Large N Extended
 Supersymmetry (II)",
 Phys.~Lett. {\bf B369} (1996) 262-268,
 hep-th/9510151;
  \bibitem{GL_Klein}
  G.~D.~Landweber:
  {\em Representation rings of Lie superalgebras},
  K-Theory 36 (2005), no. 1--2, 115--168;
 \endRefs

\end{document}